%% file: main-arxiv.tex
\documentclass[sigconf]{acmart}

\renewcommand\footnotetextcopyrightpermission[1]{} 
\usepackage{amsfonts}
\usepackage[linesnumbered,ruled,vlined,resetcount]{algorithm2e}
\makeatletter
\newcommand{\removelatexerror}{\let\@latex@error\@gobble}
\makeatother

\usepackage{multirow}
\usepackage{listings}
\usepackage{paralist}
\usepackage{stfloats}
\usepackage[caption=false,font=footnotesize]{subfig}
\usepackage{threeparttable}
\usepackage{wasysym}

\input{notations}

\AtBeginDocument{%
  }

\acmConference[CCS '25]{Proceedings of the 2025 ACM SIGSAC Conference on Computer and Communications Security}{October 13--17, 2025}{Taipei, Taiwan}

\begin{document}

\title{\sysname{}: Accurate Learning-based Static Disassembly with Attentions}

\author{Peicheng Wang}
\authornote{Both authors contributed equally to this research.}
\orcid{0009-0009-3024-6371}
\affiliation{%
  \institution{Xidian University}
  \city{Xi'an}
  \state{Shaanxi}
  \country{China}
}
\email{732176028@qq.com}

\author{Monika Santra}
\authornotemark[1]
\orcid{0000-0001-6219-6545}
\affiliation{%
  \institution{The Pennsylvania State University}
  \city{University Park}
  \state{PA}
  \country{USA}}
\email{monikas@psu.edu}

\author{Mingyu Liu}
\orcid{0009-0004-8234-177X}
\affiliation{%
  \institution{Xidian University}
  \city{Xi'an}
  \state{Shaanxi}
  \country{China}
}
\email{23151214197@stu.xidian.edu.cn}

\author{Cong Sun}
\authornote{Co-corresponding authors.}
\orcid{0000-0001-9116-2694}
\affiliation{%
 \institution{Xidian University}
 \city{Xi'an}
  \state{Shaanxi}
 \country{China}
}
\email{suncong@xidian.edu.cn}

\author{Dongrui Zeng}
\orcid{0000-0003-0032-2571}
\affiliation{%
  \institution{Palo Alto Networks, Inc.}
  \city{Santa Clara}
  \state{CA}
  \country{USA}}
\email{dzeng@paloaltonetworks.com}

\author{Gang Tan}
\authornotemark[2]
\orcid{0000-0001-6109-6091}
\affiliation{%
  \institution{The Pennsylvania State University}
  \city{University Park}
  \state{PA}
  \country{USA}}
\email{gtan@psu.edu}

\renewcommand{\shortauthors}{Peicheng Wang et al.}

\begin{abstract}
\input{./abstract}
\end{abstract}



\maketitle

\input{./intro}

\input{./background}

\input{./design}

\input{./implementation}

\input{./eval}

\input{./conclusion}

\bibliographystyle{ACM-Reference-Format}
\balance
\bibliography{mybib}

\input{./appendix}

\end{document}

%% file: notations.tex
\newcommand{\bl}[1]{\textbf{#1}}

\newcommand{\BRel}{\textit{BRel}}

\newcommand{\iVTR}{\text{iVTR}}

\newcommand{\bVTR}{\text{bVTR}}

\newcommand{\OL}[1]{\texttt{O{#1}}}
\newcommand{\ida}{IDA~Pro}
\newcommand{\ghidra}{Ghidra}
\newcommand{\ninja}{Binary Ninja}

\newcommand{\sysname}{\textsc{Disa}}
\newcommand{\xda}{XDA}
\newcommand{\bpa}{BPA}
\newcommand{\bpaDisa}{BPA$_\text{\sysname{}}$}
\newcommand{\bpaDwarf}{BPA$_\text{DWARF}$}
\newcommand{\birnn}{biRNN}
\newcommand{\deepdi}{\textsc{DeepDi}}
\newcommand{\desync}{desync-cc}
\newcommand{\ollvmfull}{Obfuscator-LLVM}
\newcommand{\ollvm}{\textsf{ollvm}}

\newcommand{\disaModel}[1]{\sysname-{#1}}
\newcommand{\xdaModel}[1]{\xda-{#1}}
\newcommand{\birnnModel}[1]{\birnn-{#1}}

\newcommand{\SPall}{SP}
\newcommand{\SNall}{SN}
\newcommand{\SOall}{SO}
\newcommand{\SRall}{SR}
\newcommand{\SP}[2]{\SPall$_\text{#1}^\text{#2}$}
\newcommand{\SN}[2]{\SNall$_\text{#1}^\text{#2}$}
\newcommand{\SO}[2]{\SOall$_\text{#1}^\text{#2}$}
\newcommand{\SR}[2]{\SRall$_\text{#1}^\text{#2}$}
\newcommand{\SNsub}[3]{\SNall$_\text{#1}^\text{#2}$-\OL{#3}}

\newcommand{\task}[1]{\textsl{T}$_\text{#1}$}

%% file: abstract.tex
For reverse engineering related security domains, such as vulnerability detection, malware analysis, and binary hardening, disassembly is crucial yet challenging.
The fundamental challenge of disassembly is to identify instruction and function boundaries. Classic approaches rely on file-format assumptions and architecture-specific heuristics to guess the boundaries, resulting in incomplete and incorrect disassembly, especially when the binary is obfuscated. Recent advancements of disassembly have demonstrated that deep learning can improve both the accuracy and efficiency of disassembly.
In this paper, we propose \sysname{}, a new learning-based disassembly approach that uses the information of superset instructions over the multi-head self-attention to learn the instructions' correlations, thus being able to infer function entry-points and instruction boundaries.
\sysname{} can further identify instructions relevant to memory block boundaries to facilitate an advanced block-memory model based value-set analysis for an accurate control flow graph (CFG) generation.
Our experiments show that \sysname{} outperforms prior deep-learning disassembly approaches in function entry-point identification, especially achieving 9.1\% and 13.2\% F1-score improvement on binaries respectively obfuscated by the disassembly desynchronization technique and popular source-level obfuscator.
By achieving an 18.5\% improvement in the memory block precision, \sysname{} generates more accurate CFGs with a 4.4\% reduction in Average Indirect Call
Targets (AICT) compared with the state-of-the-art heuristic-based approach.

%% file: intro.tex
\section{Introduction}

Disassembly is a fundamental step in analyzing, rewriting, and hardening binary programs.
The static disassembly procedure comprises different sub-tasks \cite{DBLP:conf/uss/AndriesseCVSB16, DBLP:conf/sp/PangYCKPMX21}. In these sub-tasks, identifying the function and instruction boundaries are the most fundamental tasks and prerequisites of control flow graph (CFG) construction, function signature identification, and most of the binary static analyses.
Popular static disassemblers and binary analysis tools \cite{objdump, ida-pro, ghidra, radare, binary-ninja, hopper, DBLP:conf/cav/BrumleyJAS11, Bao2014, DBLP:conf/uss/WangWW15, shoshitaishvili2016state, DBLP:conf/paste/BernatM11, b2r2}
can use the \emph{linear sweep} or the \emph{recursive traversal} algorithm combined with various architecture-specific heuristics to disassemble binaries \cite{DBLP:conf/wcre/SchwarzDA02, DBLP:conf/uss/AndriesseCVSB16}.
However, accurate disassembly is difficult.
The compiler settings and the complexity of instruction set architectures (ISAs) pose challenges to disassembly.
Moreover, even the most advanced disassembly efforts experience difficulties from missing symbols and high-level data types, embedded data in code, and especially obfuscation. Typical frameworks (e.g., \cite{ida-pro, ghidra}) place emphasis on precise disassembly and can miss a considerable set of function boundaries when the instructions of obfuscated binaries are improperly disassembled.

Machine-learning-based disassembly involves locating function entries \cite{Bao2014, DBLP:conf/uss/ShinSM15}, recovering debugging information \cite{DBLP:conf/ccs/HeITRV18}, differentiating code from data \cite{DBLP:conf/pkdd/WartellZHKT11, benkraouda2024you}, and identifying assembly instructions \cite{DBLP:conf/ndss/PeiGWYJ21, DBLP:conf/uss/YuQHY22}.
Identifying function boundaries is challenging for stripped binaries due to the lack of symbols.
Existing efforts use the weighted tree structure to learn the signatures of function starts \cite{Bao2014} or the recurrent neural network (RNN) model to predict the function boundary \cite{DBLP:conf/uss/ShinSM15}.
Despite the systematic evaluation \cite{DBLP:conf/acsac/KooPK21} demonstrating that such learning-based function identifications do not necessarily outperform the rule-based approaches \cite{ida-pro, ghidra, DBLP:conf/eurosp/AndriesseSB17}, the deep learning technique has an advantage in unifying the function and instruction boundary recovery. For example, the bidirectional RNN model \cite{DBLP:conf/uss/ShinSM15} has been extended by \cite{DBLP:conf/ndss/PeiGWYJ21} to identify the instruction boundaries, and \xda{} \cite{DBLP:conf/ndss/PeiGWYJ21} uses the BERT model \cite{kenton2019bert} to solve the two disassembly tasks by transferring the knowledge of code bytes' dependencies.

The deep-learning based disassembly approaches \cite{DBLP:conf/uss/ShinSM15, DBLP:conf/ndss/PeiGWYJ21, DBLP:conf/uss/YuQHY22} are proven to be more robust than the traditional disassembly frameworks \cite{ida-pro, ghidra}.
The deep-learning based approaches are further classified into two categories:
\begin{compactenum}[(1)]
  \item \emph{Code-byte classification} \cite{DBLP:conf/uss/ShinSM15, DBLP:conf/ndss/PeiGWYJ21} differentiates the function- or instruction-boundary bytes from the non-boundary bytes. The deep models of these classifications, e.g., bidirectional RNN \cite{DBLP:conf/uss/ShinSM15} and BERT \cite{DBLP:conf/ndss/PeiGWYJ21}, learn the byte-level correlations completely from the raw code bytes and thus are adaptive to different ISAs.

  \item \emph{Superset-instruction classification} \cite{DBLP:conf/uss/YuQHY22} identifies  true (or function-entry) instructions. This approach abstracts high-level features from a set of possible instructions using superset disassembly~\cite{DBLP:conf/ndss/BaumanLH18} and uses the relational graph convolutional network (R-GCN) model to classify instructions.
\end{compactenum}
\noindent These approaches have their respective merit in accuracy or efficiency. However, none of them was designed with the consideration of facilitating an accurate binary-level CFG generation, where the major challenge is to resolve the precise indirect call target.

\textbf{Goals}.
The code-byte classification approach can avoid the superset disassembly procedure. However, using superset disassembly as a feature-preparation step can enrich the features of the deep model with high-level knowledge. Moreover, superset disassembly relies on decoding and thus excludes erroneously decoded bytes from being predicted by the deep model. Such unnecessary predictions cannot be avoided in the code-byte classifications. On the other hand, the code-byte classification models are more straightforward to develop than the R-GCN model~\cite{DBLP:conf/uss/YuQHY22}, which requires complicated static analysis to extract typed edges between instructions, build the instruction flow graph, and adapt the graph representation to the lightweight deep model. Besides, the deep models' obfuscation resilience capability is inadequately evaluated, mainly over the binaries obfuscated with Hikari \cite{hikari}, i.e., a deprecated descendant of \ollvmfull{}~\cite{ieeespro2015-JunodRWM, ollvm-repo}, and limited to individual obfuscation options.

Therefore, our first research objective is to develop a new deep-learning based disassembly framework that takes the best of both sides to use the high-level knowledge of superset instructions and rely on simple features to learn instruction correlations.
The new deep model should apply to multiple disassembly tasks and be resilient to more obfuscation techniques, e.g.,  disassembly desynchronization~\cite{DBLP:conf/wcre/KargenHWEHS22, desync-repo} and advanced source-level obfuscation~\cite{tigress}.

Beyond identifying function entry-points and assembly instructions,
control flow graph (CFG) recovery is another crucial step of disassembly, which is challenging due to the existence of indirect calls. Reversing binary-level constructs to source-level data types can help resolve the targets of indirect calls and is also important for various binary analyses \cite{DBLP:conf/ccs/HeITRV18, DBLP:conf/dimva/MaierGWR19, DBLP:conf/sigsoft/PeiGBCYWUYRJ21}. In particular, reversed memory boundary information is critical for CFG recovery through a binary-level points-to analysis \cite{DBLP:conf/ndss/KimSZT21, DBLP:conf/cc/KimZ0T22}. Moreover,  memory boundary information is beneficial for hardening and securing binary code, e.g., vulnerability identification and mitigation \cite{DBLP:conf/usenix/SlowinskiSB12, DBLP:conf/sigsoft/WangXLLLQLL19, DBLP:conf/ispec/VaidyaKJ21} and memory sanitization for fuzzing \cite{DBLP:conf/uss/ChenSJL0DW023}.
However, existing approaches for recovering the boundary information rely on conservative heuristics, which can cause inaccurate downstream analysis.
Thus, our second research objective is to employ deep learning to recover memory boundary information. Towards this objective, our key observation is that boundary information can be inferred by a value-tracking analysis over boundary-related instructions; further, classifying boundary-related instructions aligns well with deep learning. In this way, CFG recovery is reduced to the problem of detecting boundary-related instructions.

\textbf{Method}.
In this work, we propose a new learning-based disassembly approach called \sysname{}. It targets three disassembly tasks: \task{1}) function entry-point identification, \task{2}) true instruction identification, and \task{3}) memory block boundary identification for improving CFG recovery.
The three tasks are conducted independently of each other using different deep models that share a homogeneous model architecture.
The first step of \sysname{} is the superset disassembly, which decodes the bytes at each possible code address to derive the information of the overlapping superset of potential instructions.
\sysname{} uses the information of decoded instructions over the multi-head self-attention to learn the instructions' correlations in these task scenarios.
Benefiting from the high-level knowledge of superset instructions and the encoder's capability of discovering long-range dependencies, \sysname{} has an outstanding performance in disassembling binaries obfuscated by \cite{tigress, DBLP:conf/wcre/KargenHWEHS22}.

Besides identifying true instructions and function-entry instructions, \sysname{} has another capability of identifying memory block boundaries.
\sysname{}'s instruction representation and embedding are appropriate for classifying instructions instead of instruction operands. As a disassembly framework that predicts  superset instructions,
\sysname{}'s deep model first classifies memory-accessing superset instructions to identify instructions that manipulate block-boundary data elements, i.e., block-boundary-related (\BRel) instructions.
Then, \sysname{} uses a post-classification value-tracking procedure to translate the decision of \BRel{} instructions into valid memory block boundaries.
Our design of the bidirectional value-tracking analysis makes deep-learning based disassembly approaches applicable to memory block boundary identification, which is critical to downstream applications such as  binary-level points-to analysis and CFG recovery.
By integrating the results of the tasks into the block-memory model based points-to analysis (BPA)~\cite{DBLP:conf/ndss/KimSZT21}, \sysname{} makes a novel combination of deep models and static analysis to resolve indirect call targets and construct accurate, assembly-level CFGs.
The self-attention mechanism of \sysname{}'s deep model can efficiently learn the code features and ease the labor of static analysis as used by related approaches \cite{DBLP:conf/uss/YuQHY22, DBLP:conf/ccs/HeITRV18, DBLP:conf/dimva/MaierGWR19}.

\textbf{Evaluation}.
Our evaluations on tasks \task{1} and \task{2} are focused on the accuracy, robustness, and efficiency of \sysname{} compared with the state-of-the-art learning-based disassembly approaches \cite{DBLP:conf/uss/ShinSM15, DBLP:conf/ndss/PeiGWYJ21, DBLP:conf/uss/YuQHY22} and traditional frameworks \cite{ida-pro, ghidra}.
After evaluating the accuracy of the best deep models we can obtain, we demonstrate \sysname{}'s generalizability over real-world programs, using binaries built with complex optimization options discovered by BinTuner~\cite{DBLP:conf/pldi/RenHMLL21}, and the binaries obfuscated with Obfuscator-LLVM, \desync{}, tigress, VMProtect, and Themida.
Our investigations use more obfuscators and more intensive obfuscation options than previous work~\cite{DBLP:conf/ndss/PeiGWYJ21, DBLP:conf/uss/YuQHY22}, combined with different compiler optimization levels, demonstrating \sysname{}'s efficacy in real-world obfuscation.
Besides, we also build more fine-grained deep models with x64-ELF binaries to investigate their capability to generalize to obfuscated binaries. Efficiency evaluation on obfuscated binaries shows that \sysname{} outperforms \xda{} and \ghidra{} and is competitive with other approaches. For evaluating task \task{3}, we demonstrate that the \sysname{}-based improvement of BPA outperforms the original BPA, approximating the performance of the reference BPA using block boundary ground truths for the points-to analysis. Additionally, the ablation study shows that BPA with \sysname{} is more effective in recovering the CFG when all \sysname{} tasks, i.e., \task{1}, \task{2}, and \task{3}, are included.

\textbf{Contributions}. We make the following contributions.
\begin{compactenum}[(1)]
  \item \sysname{} outperforms SOTA classification approaches \cite{DBLP:conf/uss/ShinSM15, DBLP:conf/ndss/PeiGWYJ21, DBLP:conf/uss/YuQHY22} on the robustness in disassembling binaries obfuscated with the disassembly desynchronization technique~\cite{DBLP:conf/wcre/KargenHWEHS22} and identifying function entries of binaries obfuscated with advanced source-level obfuscation technique~\cite{tigress}. Specifically, \sysname{} improves the SOTA function entry-point identifications by 9.1\% against \cite{DBLP:conf/wcre/KargenHWEHS22} and 13.2\% against \cite{tigress}.

  \item \sysname{} is the first disassembly approach that can identify memory block boundaries at the binary level. Unlike the learning-based binary data type inference approaches, \sysname{} simplifies the prediction of complicated instruction operands into the determination of block-boundary-related instructions with a value-tracking analysis. \sysname{} excels in memory block boundary identification, outperforming BPA by 18.5\% in precision, 11.6\% in recall, and 16.1\% in F1-score. In particular, when it comes to local memory block partitioning, \sysname{} surpasses \bpa{} with 27\%, 21.3\%, and 24\% in precision, recall, and F1-score, respectively.

  \item We improve \bpa{} \cite{DBLP:conf/ndss/KimSZT21} by integrating \sysname{}. By replacing \bpa{}'s heuristics-based memory blocks with \sysname{}'s more precise memory blocks, we achieve a 4.4\% reduction in AICT and demonstrate the potential to offer more precise CFGs, better suited for downstream security applications. As a result, \sysname{} and \sysname{}-facilitated BPA achieve the assembly-level accurate CFG construction from the raw binary.
      \sysname{}'s implementation is publicly available at \url{https://github.com/peicwang/Disa}.
\end{compactenum}

%% file: background.tex
\section{Background}\label{sec:background}

\subsection{Scope of \sysname{}}
Locating function entry-points, identifying assembly instructions, and recovering binary-level CFG are the three primary steps of static binary disassembly.
\sysname{} is designed to focus on such primary steps.
These primary steps are relied upon by more advanced reverse-engineering steps \cite{DBLP:conf/uss/AndriesseCVSB16, DBLP:conf/sp/PangYCKPMX21} such as symbolization and recovering function signatures.
There are two major types of CFG recovery methodologies: signature matching and points-to analysis. We craft \sysname{} to help improve the block-memory model based points-to analysis (\bpa{}) \cite{DBLP:conf/ndss/KimSZT21} by identifying memory-block boundary related instructions, since \bpa{} is the most practical points-to analysis for CFG recovery in terms of accuracy and efficiency. More background on CFG recovery can be found in Section~\ref{subsec:cfg-recovery}.
Moreover, memory blocks can be of different granularity. For example, one can design a points-to analysis to accommodate compound memory blocks (i.e., a group of memory blocks as one super block). In theory, \sysname{} can be applied to generate compound memory blocks, as long as the ground-truth data for training contains compound memory blocks. However, in this paper, we limit our scope to memory blocks corresponding to \bpa{}'s definition. In other words, one memory block should be one continuous range of memory locations, mapping to one source-level compound data structure. Thus, memory blocks in our discussion can be perfectly identified by source code or debugging information.
The \sysname{} improvement of BPA (i.e., \bpaDisa{} in Section~\ref{sec:implementation}) is compared in CFG recovery effectiveness with the original BPA and a reference version of BPA (i.e., \bpaDwarf{} in Section~\ref{subsec:baseline}) which uses the DWARF-based memory block ground truths to resolve indirect branches. Comparing \bpaDisa with signature matching based CFG recovery is out of our scope since they do not rely on memory blocks and our comparison is meant to show the effectiveness of \sysname{}'s memory block prediction.

\sysname{} focuses on static disassembly on stripped x86/x64 binaries. Moreover, \sysname{}'s function entry and instruction identifications accommodate binaries obfuscated by several source-level obfuscators, i.e., Obfuscator-LLVM, desync-cc, and tigress. Especially, the disassembly desynchronization implemented by~\cite{DBLP:conf/wcre/KargenHWEHS22}
uses opaque predicates to trick disassemblers into decoding the machine-code sequence at wrong offsets, which potentially yields valid albeit meaningless assembly code. Such obfuscation features are captured by  deep-learning based disassemblers like \sysname{} but pose a significant obstacle to traditional frameworks.
The static disassemblers like \sysname{} can hardly disassemble dynamically generated code or runtime code decryption. Several assembly-level obfuscators, e.g., VMProtect~\cite{vmprotect} and Themida~\cite{themida}, intensively rely on these techniques to hide code. Our evaluation in Section~\ref{sec:evaluation} reports the static disassemblers' limited capability over a large part of the runtime-reached code produced by these obfuscators.
Since the binary-level points-to analyses, e.g., \bpa{}, Value-Set Analysis, and \sysname{}'s post-classification value tracking, have yet accommodated obfuscated binaries, \sysname{} generates assembly-level CFG on obfuscated binaries without resolving indirect edges, even though \sysname{}'s deep model has a good potential to predict the \BRel{} instructions of these binaries.
Incorporating dynamic techniques, e.g., \cite{Bonfante2015, DBLP:conf/sp/BardinDM17, DBLP:conf/cc/FedericoPA17} is left for future work.

\subsection{Control Flow Graph Recovery}\label{subsec:cfg-recovery}
Control flow graph (CFG) recovery from binary is one important step of disassembly, where the most challenging task is to decide the indirect call edges.
However, the SOTA superset-disassembly based CFG recovery either omits indirect calls~\cite{DBLP:conf/uss/KruegelRVV04, DBLP:conf/pakdd/WartellZHK14, DBLP:conf/icse/MillerKSZZL19, DBLP:conf/cases/KhadraSK16, DBLP:conf/uss/Flores-MontoyaS20}, or converts the indirect branch resolving into a rewriting-based mapping lookup~\cite{DBLP:conf/ndss/BaumanLH18}.
During runtime, indirect calls can target different functions at different times. Thus, our task is to find a set of functions as the possible targets of each indirect callsite.
Typical approaches include points-to analysis (e.g., Value-Set Analysis (VSA) \cite{DBLP:conf/cc/BalakrishnanR04} and \bpa{} \cite{DBLP:conf/ndss/KimSZT21}) and signature matching methods (e.g., TypeArmor \cite{DBLP:conf/sp/VeenGCPCRBHAG16} and $\tau$CFI \cite{DBLP:conf/raid/MunteanFTLGE18}). Among these techniques, points-to analysis can achieve the best accuracy of CFG recovery from stripped binaries. Note that inferring data types and data structures can further improve the accuracy of indirect call edges \cite{DBLP:journals/csur/CaballeroL16, DBLP:conf/sp/ZhangYYTLKAZ21, DBLP:conf/sp/LinG21}. In addition to classic program-analysis based approaches, recent efforts \cite{zhu2023callee} \cite{attncall} have used deep neural networks to predict indirect call edges by transferring knowledge of direct calls' contexts. The scope of usage of these works differs from our general-purpose disassembly framework.

\subsection{Memory Block Generation of \bpa{}}

Points-to analysis is fundamental for recovering CFGs of stripped binaries. VSA and VSA variations, however, are too expensive to scale. In contrast, \bpa{}~\cite{DBLP:conf/ndss/KimSZT21} implements points-to analysis based on a \emph{block memory model}, the key to achieving both scalability and precision. In this memory model, the  memory is modeled as a set of disjoint \emph{memory blocks}. Each block is comprised of a logically cohesive set of memory locations. For example, the stack of a function can be divided into a set of memory blocks, with one block for each local variable.
All locations in a block are treated as equivalent during points-to analysis; that is,  any read/write to a location in a block is treated as operating on all locations in the block. By using memory blocks, \bpa{} avoids reasoning about pointer arithmetics; specifically,  the block memory model assumes that it is not possible to make a pointer to one block point to a different block via pointer arithmetics. \bpa{} takes advantage of this to achieve much better scalability than VSA.

The accuracy of memory block boundaries, however, becomes one deciding factor for \bpa{}'s points-to analysis. Coarse-grained boundaries are more scalable, but result in less precision.  An accurate memory block is defined as a set of memory locations that correspond to some source-level compound data structure. \bpa{} uses heuristics to generate memory blocks, with the principle that any pointer arithmetic needs to stay within a block. More details of \bpa{}'s memory block generation can be found in the paper~\cite{DBLP:conf/ndss/KimSZT21}.

\bpa{}'s manually designed heuristics can be either too conservative or too aggressive, causing  many false positives or false negatives. Therefore, \sysname{}'s task \task{3} is designed to obtain accurate memory block boundaries by deep learning, so that the block-memory model based points-to analysis such as \bpa{} can achieve better accuracy in recovering CFGs. As a result,
\sysname{} is the first work that extends deep learning from function entry and instruction identification to accurate CFG construction,
while other deep-learning-based approaches \cite{DBLP:conf/uss/ShinSM15, DBLP:conf/ndss/PeiGWYJ21, DBLP:conf/uss/YuQHY22} only focus on the first two tasks of \sysname{}. 

%% file: design.tex
\section{Design of \sysname{}}\label{sec:design}

\begin{figure*}[t]\centering
  \includegraphics[width=0.7\textwidth]{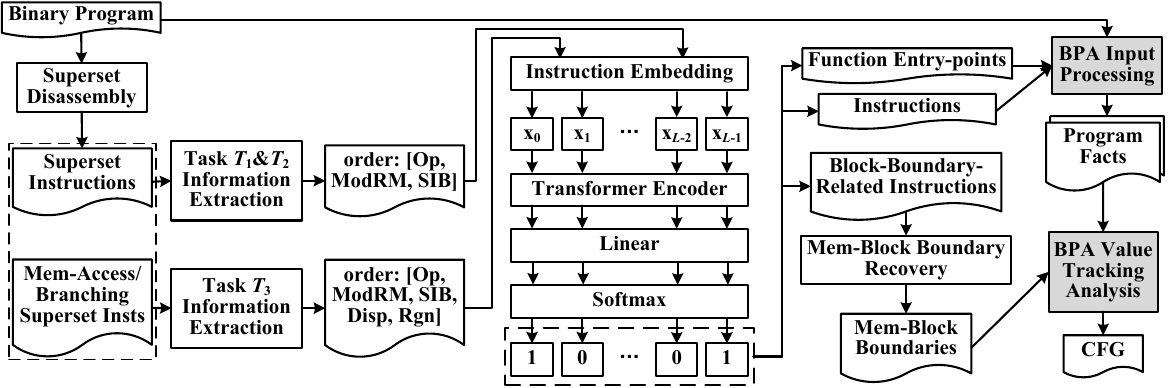}%
\vspace{-2ex}
  \caption{Workflow of \sysname{}}
  \label{fig:disa-workflow}
\vspace{-2ex}
\end{figure*}

The workflow of \sysname{} is presented in Fig.~\ref{fig:disa-workflow}. \sysname{} has three disassembly tasks: function entry-point identification (\task{1}), true instruction identification (\task{2}), and memory block boundary identification (\task{3}).
For each task, \sysname{} adopts a deep neural network model, which makes binary classification on instructions resulting from superset disassembly.
The model of task \task{1} classifies function entry-point instructions; the model of task \task{2} determines whether a superset-disassembly instruction is a true instruction; and the model of task \task{3} identifies \emph{block-boundary-related (\BRel{}) instructions}.
{A \emph{block-boundary-related instruction} is a true instruction that accesses a memory block boundary address. An instruction may access multiple memory addresses in different executions, e.g., an instruction in a loop. In that case, the instruction is \BRel{} instruction if at least one of these memory addresses is a block boundary address.}
Then, we use a static boundary recovery procedure to transform \task{3}'s prediction results into memory block boundaries.
As a security application, the block boundaries recovered from \task{3}'s prediction results are delivered to \bpa{} to predict the targets of indirect calls. We call the integrated system, \bpaDisa{}. In addition, the prediction results of \task{1} and \task{2} are provided to \bpaDisa{} for the integration of three \sysname{} tasks, resulting in BPA$_{\sysname{}}^{\text{T12}}$.

\sysname{} implements two decode primitives over the results of superset disassembly to extract all the instruction fields used in the learning-based disassembly tasks. Differently from the fields used by \cite{DBLP:conf/uss/YuQHY22}, we use more high-level knowledge to infer the features of the manipulated data memory addresses from the superset instructions for task \task{3}. Therefore, our decode primitive extracts the \emph{memory region} and \emph{relative displacement} information from memory-access instructions, in combination with other addressing fields to decide the \BRel{} instructions (Section~\ref{subsec:semantic-extraction}).
The key component of our deep model is the encoder structure of the transformer. For each disassembly task, the encoder's multi-head self-attention mechanism learns the relevances among the superset instructions, predicts valid instruction sequence patterns through a binary classification, and finally derives the function entries, true instructions, or \BRel{} instructions (Section~\ref{subsec:embedding}).
We then use intra-procedural value tracking analysis to recover the memory block boundaries for \bpaDisa{} from the outputs of \task{3}'s deep model (Section~\ref{subsec:boundary-recovery}).

\subsection{Extracting Instruction Information}\label{subsec:semantic-extraction}

Decoding instructions from  raw bytes at each offset of the code section of an input binary is guaranteed to obtain a superset of true instructions. This superset disassembly \cite{DBLP:conf/ndss/BaumanLH18} causes no instruction-level false negatives, thus reducing the disassembly task into an instruction classification problem on the superset.
Although the instruction superset is not essential to a deep-learning-based disassembly (e.g., one can do raw-byte classification as in XDA), we take the superset disassembly as the first step to obtain the knowledge of instructions, as performed by \cite{DBLP:conf/uss/YuQHY22, DBLP:conf/sp/YeZSAZ23}. For a code byte vector $b_{0..S-1}$, we obtain the superset $\textsl{SI}=\{ins_i\mid ins_i\leftarrow decode(b,i), 0\leq i<S\}$.
$decode(b,i)$ linearly disassembles the bytes at the offset $i$ and returns an assembly instruction or a decoding failure.
Since decoding may fail, $\textsl{SI}$ generally can have less than $S$ instructions, ordered by their offsets from the start of the code section.

Conceptually, a decoding primitive $decode(b,i)$ takes a sequence of bytes and outputs all information about the instruction encoded by the bytes. Practically, we extract only necessary information from the decoded instruction.
Representing the instruction as text tokens, e.g., \cite{DBLP:conf/ccs/LiQY21}, would cause an enormous dictionary and inefficient model training procedure. Instead,
we define two decoding primitives to extract typical fields from each instruction and map the value of these fields as integers for model training:
\begin{align*}
  decode_1(b,i) & = (\textit{Op}_i,\textit{ModRM}_i,\textit{SIB}_i) \\
  decode_2(b,i) & = (\textit{Op}_i,\textit{ModRM}_i, \textit{SIB}_i, \textit{Disp}_i, \textit{Rgn}_i)
\end{align*}
\noindent For the models of function entry and instruction identification, $decode_1$ extracts \textit{Opcode}, \textit{ModRM}, and \textit{SIB}\footnote{The \textit{ModRM} byte specifies the addressing mode and register combination for the instruction. The \textit{Scale-Index-Base (SIB)} byte follows the \textit{ModRM} byte to represents more flexible addressing memory operands.} from an instruction.
The three fields are encoded into disjoint integer ranges, ensuring each possible value of the three fields has unambiguous meaning. Thus, $decode_1$ outputs a triple of integers as the representation of the instruction.
For robustness, we do not use any instruction prefix.

\begin{figure}[t]\centering
  \includegraphics[width=0.85\columnwidth]{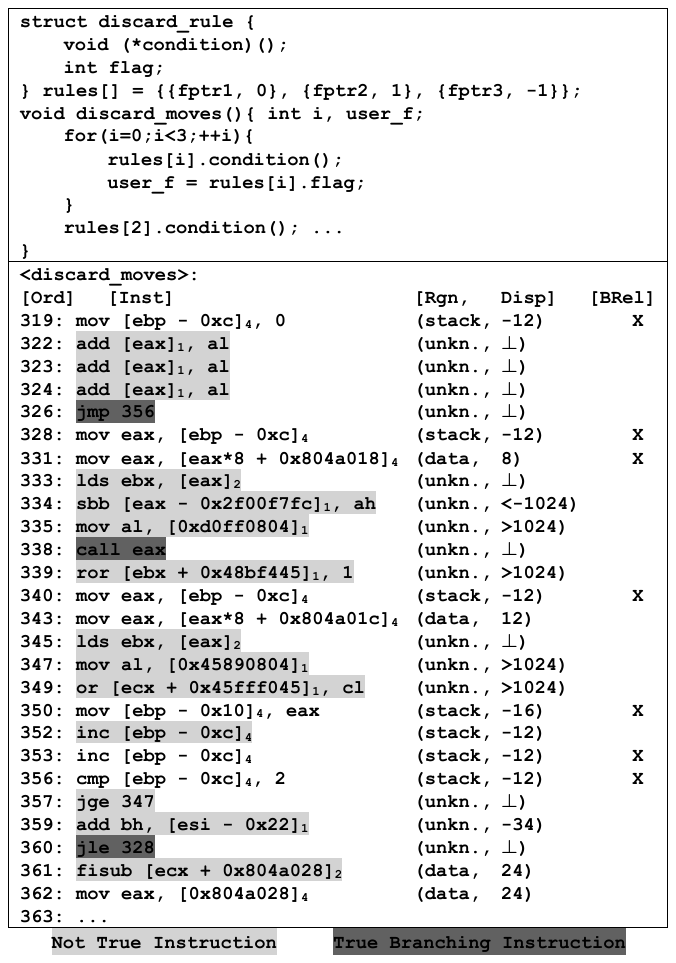}%
\vspace{-2ex}
  \caption{Example with Auxiliary Features and \BRel{} Instructions (\textsf{upper}=\textsf{lower}=1024, \textsf{.data} is [0x804a010, 0x804a02f]).
  }
  \label{fig:example}
\vspace{-4ex}
\end{figure}

Memory block boundaries are related to data operations over compound data structures.
A stack memory block boundary can be represented as an offset from the stack pointer (\verb|esp|) at the function's entry. A global memory block boundary is an offset from the start of global sections (\textsf{.data}, \textsf{.rodata}, \textsf{.bss}).
For task \task{3}, $decode_2$ extracts more information to represent the feature of memory operations on the global and stack data structures.
Firstly, \task{3} is concerned with only memory-access instructions and branching instructions. Other instructions are not decoded in task \task{3}.
Instead of applying complex analyses over instructions \cite{DBLP:conf/uss/YuQHY22}, we use a simple, per-instruction memory-access knowledge extraction.

Specifically, we derive a \emph{memory region} (\textit{Rgn}) and a \emph{relative displacement} (\textit{Disp}) as auxiliary memory access features.
An instruction's \textit{Rgn} is an integer encoding a region from
\[\{\textsf{heap}, \textsf{stack}, \textsf{rodata}, \textsf{data}, \textsf{bss}, \textsf{unknown}\},\]
indicating where the memory block manipulated by the instruction is located.
We use heuristic-based pattern matching to decide the \textit{Rgn} value of each instruction.
These patterns include specific registers, constant values, and whether the constants fall in specific global sections, which are computed by the instruction's \textit{ModRM}, \textit{SIB}, and \textit{Displacement}\footnote{\textit{Displacement} specifies a static (uncomputed) address or offset for addressing.}.
If we cannot track the region of a memory-access expression during the instruction-decoding procedure, \textit{Rgn} is labeled \textsf{unknown}.

The instruction's relative displacement \textit{Disp} is the offset from a base addressing expression in the data sections or a stack frame.
{For encoding convenience, \textit{Disp} can differ from \textit{Displacement}. For examples in Fig.~\ref{fig:example}, the \textit{Displacement} 0x804a018 of instruction 331 has the corresponding \textit{Disp} 8 because 0x804a018 falls into the \textsf{.data} section [0x804a010, 0x804a02f], while for instruction 319, the \textit{Displacement} ($-$0xc) is an offset and identical to the \textit{Disp} ($-$12).}
The representation of \textit{Disp} leads to an integer encoding of $\{\bot,$ $<$\textsf{lower}, $\textsf{lower}$, $\ldots$, \textsf{upper}, $>$\textsf{upper}$\}$, where $\bot$ means the absence of \textit{Displacement}.
[$\textsf{lower}$, \textsf{upper}] is a fixed range of consecutive integers that represent those exact relative displacements tracked by our model, e.g., from -1024 to 1024. This range is selected to not intersect with $\bot$, $<$\textsf{lower}, $>$\textsf{upper}, \textit{Opcode}, \textit{ModRM}, \textit{SIB}, and \textit{Rgn}. The branching instruction positively impacts the decision of \BRel{} instructions. To unify the instruction representation of task \task{3}, the branching instructions are properly encoded with (\textit{Rgn}, \textit{Disp})=(\textsf{unknown}, $\bot$).
As a result, $decode_2$ returns a quintuple of integers to represent each memory-access or branching superset instruction.

For an example adapted from \verb|gobmk| of SPEC2k6, Fig.~\ref{fig:example} presents the superset memory-access and branching instructions of the function \verb|discard_moves|. {Decoding failures are not presented.}
We deem that the memory-access patterns of instructions reflect the boundary features of compound data structures. The labels of \BRel{} instructions are derived from the ground truth, as described in Section~\ref{subsec:boundary-recovery}.
Task \task{3}'s model training uses these labels but does not use the information about true instructions; thus, tasks \task{3} and \task{2} are independent.
For simplicity, Fig.~\ref{fig:example} elides the results of \textit{Opcode}, \textit{ModRM}, and \textit{SIB} and includes only
(\textit{Rgn}, \textit{Disp}).
As shown in the figure, the decoded quintuples of true \BRel{} instructions differ in pattern from the block-boundary-unrelated memory-access instructions.
In the example, an instruction within \textsf{unknown} region tends to be predicted as not a block-boundary instruction. The local variables \verb|i| and \verb|user_f| respectively locate at \verb|[ebp-0xc]|$_4$ and \verb|[ebp-0x10]|$_4$. Their boundaries are memory block boundary thus the true instructions accessing them, i.e., instructions at 319, 328, 340, 350, 353, and 356, are \BRel{} instructions.
Considering the global compound data structure \verb|rules|, \verb|rules[0]| starting at 0x804a018 with (\textit{Rgn}, \textit{Disp})=(\verb|data|, 8) is a memory block boundary, and the instruction accessing the global region at 331 is a \BRel{} instruction.
In contrast, the instructions 343 and 362 respectively access \verb|rules[i].flag| and \verb|rules[2]|. The global addresses touched by these instructions are not memory block boundary because there are pointer arithmetics crossing these addresses.
Differentiating the \BRel{} instruction 331 from the block-boundary-unrelated instructions (343 and 362) with similar (\textit{Rgn}, \textit{Disp}) pairs requires more context knowledge.
Our deep model learns the correlations among the contexting quintuples of the memory-access and branching instructions to correctly predict these instructions. The knowledge of branching instructions, either true (326, 338, and 360 of Fig.~\ref{fig:example}) or not true (357 of Fig.~\ref{fig:example}), also helps classify the \BRel{} instructions.

\subsection{Instruction Embedding and Self-Attention}\label{subsec:embedding}

The instruction fields and position are both critical knowledge for disassembly. Each superset instruction's position is critical to reflect the context of the instruction in the final disassembly code. For example, a conditional branch instruction following \verb|cmp| is more likely than other kinds of instructions. Therefore, \sysname{}'s instruction embedding consists of \emph{instruction position embedding} and \emph{instruction fields embedding}.

Modeling the relevances of all $N$ instructions in the superset $\textsl{SI}$ for classification would be costly and unnecessary for disassembly. Instead, we use a smaller, constant-length input sequence with length $L$ to embed the superset instructions.
The superset instructions of a binary are separated into $\lceil N/L\rceil$ input sequences, and the last input sequence would be padded in tail with a particular padding integer.
Each element of the input sequence is indexed from 0 to $L-1$, indicating the relative position.
A learned embedding layer takes the indices as input to embed the relative instruction positions. The output indicates all the position information of the instruction sequence.
For the instruction fields embedding, the fields of each triple of \task{1} and \task{2} are respectively embedded with three learned embedding layers.
We concatenate the three embeddings to form the instruction fields embedding with dimension $d_\text{model}$ \cite{DBLP:conf/nips/VaswaniSPUJGKP17}.
Similarly, for the integer quintuple of each instruction in \task{3}, we use an embedding layer for each field and concatenate the five embeddings into an instruction fields embedding.
Adding each dimension of the position and fields embedding derives the final instruction embedding of the instruction sequence.

\sysname{} uses the encoder structure of the transformer \cite{DBLP:conf/nips/VaswaniSPUJGKP17} to classify superset instructions.
Each layer of the encoder iteratively updates the instruction embedding of the instruction sequence and derives the representations capturing the patterns of true function entries or (\BRel{}) instructions.
The output embeddings of the encoder are delivered to a linear and a softmax layer to predict the probability of the verdict on each superset instruction.
Because \sysname{}'s deep model learns the relevances among superset instructions from the binary labeling on instructions, our model can use the instructions' relevance knowledge to determine the binary classification by predicting the binary labeling patterns.
The capability of the encoder structure to learn the high-level semantic relevance of superset instructions makes \sysname{}'s model more effective and robust than the raw-byte based deep models~\cite{DBLP:conf/uss/ShinSM15, DBLP:conf/ndss/PeiGWYJ21} and the R-GCN model~\cite{DBLP:conf/uss/YuQHY22}, which requires complex prior-static analysis on task \task{2}; as a result, \sysname{} is easier to develop than the R-GCN model~\cite{DBLP:conf/uss/YuQHY22}.

\subsection{Memory Block Boundary Recovery}\label{subsec:boundary-recovery}

Task \task{3} uses a deep-learning model to identify instructions that access memory-block boundary addresses. The model's prediction results are about which instructions are boundary-related instructions and are not directly usable for systems such as \bpa{}, which requires memory block boundary information. To fill in the gap, we devise a follow-up static analysis to recover memory block boundaries from task \task{3}'s results. Next we discuss this static analysis
and after that we will elaborate on the process of generating the ground truth for \task{3}'s model training in Section~\ref{subsubsec:gt-train}.

The static analysis is an intra-procedural value tracking analysis, which we call \emph{boundary-targeted value tracking analysis} (\bVTR{}). This analysis is conducted to convert the memory operands of instructions into memory block boundaries, i.e., as offsets from a function \textit{F}$_i$'s entry value of \verb|esp| or some global region.

\begin{figure*}[!t]\centering
  \includegraphics[width=0.8\textwidth]{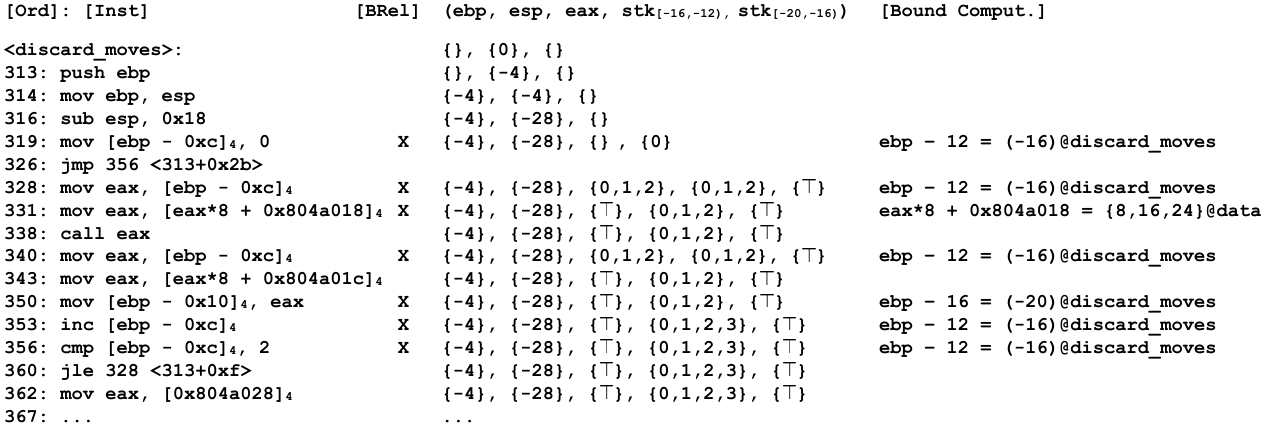}%
\vspace{-2ex}
  \caption{Value Tracking with Boundary Identification}
  \label{fig:value-track}
\vspace{-3ex}
\end{figure*}

The primary logic of \bVTR{} revolves around the value set analysis, as sketched in Algorithm \ref{alg:value_tracking_analysis2}.
The \bVTR{} algorithm takes as input the binary file and the predicted \BRel{} instruction addresses provided by DISA \task{3}. The output of the analysis is a set of pairs $(\textit{Rgn}, \textit{MemBlocks})$, where \textit{Rgn}  is a global region such as \textsf{.rodata}, \textsf{.bss}, \textsf{.data}, or a function name for the function's stack,
and \textit{MemBlocks} is the memory block boundaries of \textit{Rgn}.
In order to determine the final memory block boundaries, the \bVTR{} invokes the \texttt{FuncWiseVSA} procedure for each function in the binary to calculate instruction-level \texttt{ValueSets}. In general, a value set tells the set of possible values of a register or a stack slot. For a pointer to the stack, it tracks the set of possible offests to the inital esp value at the function entry. For example, after instruction 313 (a push instruction) in
Fig. \ref{fig:value-track}, esp's offset to the inital esp value is -4; after instruction 314, ebp's offset to the inital esp value is also -4.

\removelatexerror
\begin{algorithm}[H]\small
\setcounter{AlgoLine}{0} 
\caption{Boundary-Targeted Value Tracking}
\label{alg:value_tracking_analysis2}
\KwIn{Binary file, DISA \task{3} predictions}
\KwOut{(Rgn, MemBlocks)}

\ForEach{function in the binary}{
    \tcp{\scriptsize \textcolor{blue}{Conducts VSA and returns value sets for each instruction}}
    ValueSets $\leftarrow$ FuncWiseVSA(disassmFunc)\;

    \tcp{\scriptsize \textcolor{blue}{Returns region-wise memory blocks}}
    $(Rgn, MemBlocks) \leftarrow \text{GetMemBlocksFromDISA}(\text{ValueSets}, \text{\task{3} predictions})$\;
}
\end{algorithm}

After the function-level value-set analysis, {\small \texttt{GetMemBlocksFromDISA}} processes each predicted \BRel{} instruction by referencing and simplifying the memory operand using its computed \texttt{ValueSets} result to derive the final \textit{Memblocks}. Since only those boundary-related instructions are supposed to access memory block boundaries, using those operands' value set results would produce the boundaries of memory blocks.
For example, at instruction 319 in Fig. \ref{fig:value-track}, we derive the value of operand \verb|ebp-12| with \verb|ebp|'s current value set and obtain \verb|(-16)@discard_moves| as a valid boundary.
Note that our algorithm tracks the values in global sections, though we ignore them in Fig.~\ref{fig:value-track} for simplicity.

{\small \texttt{GetMemBlocksFromDISA}} may derive multiple block boundaries from one instruction; e.g., on instruction 331, we obtain three boundaries \verb|{8,16,24}@data| due to three possible values of \verb|eax|. Our instruction-level prediction prohibits \sysname{} from differentiating these boundaries if only part of them are real boundaries: if one instruction touches both true and false block boundaries, our algorithm labels such an instruction as \BRel{}, and our system could introduce false-positive block boundaries.
For example, even though \verb|24@data| is not a true memory boundary due to the \BRel{} label at 362 and \sysname{}'s \task{3} model can make the correct prediction, such global address will still be decided as block boundary by \bVTR{} over instruction 331.
Besides, due to the potential imprecision of the value tracking, e.g., heap values are ignored, Task \task{3}'s prediction may result in having an offset on the \textsf{unknown} region.  In that case, we deem this offset not a memory block boundary.

\subsubsection{Ground Truth Collection for \task{3} Model Training}\label{subsubsec:gt-train}
The ground truth collection for \task{3} consists of two parts. The first involves extracting memory block boundaries from DWARF information. The second entails using these extracted boundaries to obtain ground truths for training \task{3}'s deep model, i.e., those \BRel{} instructions.

\textbf{Extracting Block Boundaries.}
DWARF \cite{dwarf} is a debugging information format that encodes source-code information in binary code.
The DWARF debugging information is a series of debugging information entries (DIEs)  such as data types and locations of variables, arrays, and other complex data structures. Since debugging information is needed by generic debuggers, compilers need to produce debugging information with high precision, even if optimizations are turned on. The type information and location information inside the debugging information can help determine the boundaries of memory blocks. For example, if the debugging information tells that the first element of an array of 10 bytes resides in memory address 1000, we can infer that the memory block that represents the array has the range $[1000,1009]$. Assuming that pointer arithmetic should not cross the boundary of source-level data structures, memory blocks extracted from debugging information can be treated as ground truth.

We extract debugging information entries (DIEs) for variables, functions, and location tags, equivalent to DWARF representations of \texttt{DW\_TAG\_Variable}, \texttt{DW\_TAG\_Subprogram}, and \texttt{DW\_AT\_location}, respectively. Subsequently, these entries are processed to yield output formatted as (\textit{Rgn}, \textit{offsets}).
The \textit{offsets} denote the list of computed offsets from the beginning of a global region or, for the stack, from the initial \texttt{esp} of a function.
Notably, DWARF provides stack location information in various forms, such as base registers like \texttt{ebp}/\texttt{esp} or call frame base, coupled with a displacement value. Changes in the relationship between these base registers and the function's starting \texttt{esp} may occur based on different optimizations. To address this, we perform a function prologue analysis to standardize the debugging information into our format, ensuring alignment of every offset with the initial \texttt{esp}.

\begin{figure}[!t]
\centering
\includegraphics[width=0.85\linewidth]{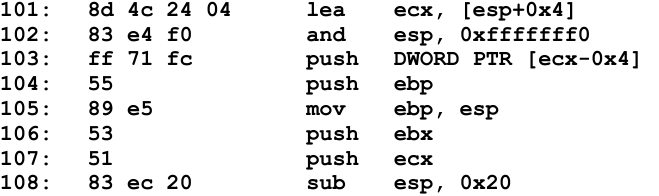}
\vspace{-2ex}
\caption{Sample Function Prologue}\label{fig:func_prolog}
\vspace{-2ex}
\end{figure}

\begin{figure}[!t]
\centering
\includegraphics[width=0.75\linewidth]{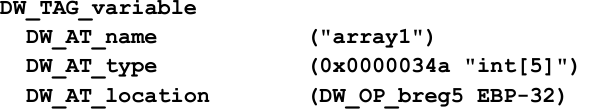}
\vspace{-2ex}
\caption{Sample DWARF for a Variable}\label{fig:sample_dwarf}
\vspace{-3ex}
\end{figure}

For example, a sample function prologue is presented in Fig. \ref{fig:func_prolog}. A close inspection reveals the presence of two push operations at addresses 103 and 104, preceding the initialization of \verb|ebp| with the value of \verb|esp|. As a result, the distance between the initial \verb|esp| and the \verb|ebp| is determined to be 8 bytes, with each push operation consuming 4 bytes. Fig. \ref{fig:sample_dwarf} displays the DWARF information representing a variable denoted as \texttt{array1}. For brevity, other attributes of the variables are elided. The variable's location is indicated using the DWARF tag \texttt{DW\_AT\_location}, offering the positional data as \texttt{DW\_OP\_breg5 EBP-32}. In the DWARF format for 32-bit x86 binaries, \texttt{DW\_OP\_breg5} symbolizes \verb|ebp|. The location data signifies the \verb|ebp| base registers and an offset of $-32$. Since our output reflects the offset from the initial \verb|esp|, adjusting the $-32$ by 8 allows us to ascertain that this memory block commences from an offset of $-40$ from the initial \verb|esp| specifically for this function.

\textbf{Identify Memory-Access Instructions.}
To accurately establish the required ground truth for DISA \task{3}, we perform another value-tracking analysis, known as the \emph{instruction-targeted value tracking analysis} (\iVTR{}). Similar to \bVTR{}, the key focus of \iVTR{} is also the value set analysis. We extract the value sets for each assembly instruction and then compare them with the raw memory block offsets obtained with DWARF information.
This approach enables us to pinpoint the memory addresses that accesse memory block boundaries.

\begin{algorithm}[H]\small
\caption{Instruction-Targeted Value Tracking}
\label{alg:value_tracking_analysis}
\KwIn{Binary file}
\KwOut{(\textit{Rgn, instAddrs})}

\tcp{\scriptsize \textcolor{blue}{Extracts the raw memory blocks from DWARF}}
$(Rgn, offsets) \leftarrow \text{ExtBlkBndDWARF}(\text{binary})$\;

\ForEach{function in the binary}{
    \tcp{\scriptsize \textcolor{blue}{Conducts VSA and returns value sets for each instruction}}
    ValueSets $\leftarrow$ FuncWiseVSA(funcName, binary)\;

    \tcp{\scriptsize \textcolor{blue}{Returns DISA task3 formatted ground truth}}
    $(Rgn, instAddrs) \leftarrow \text{IdInstrTouchMem}(\text{ValueSets}, (Rgn, offsets))$\;
}
\end{algorithm}

The \iVTR{} algorithm in Algorithm \ref{alg:value_tracking_analysis} takes the binary itself as input and produces output in the form of $(\textit{Rgn}, \textit{instAddrs})$.
\textit{instAddrs} consists of the list of instruction addresses that interact with memory block boundaries, we label these memory-access instructions as \BRel{}. In the first step, \iVTR{} calls the \texttt{ExtBlkBndDWARF} function, which processes the binary and returns the raw offsets in the format of $(\textit{Rgn}, \textit{offsets})$.
Subsequently, for each identified function in the binary, we execute \texttt{FuncWiseVSA} to obtain the instruction-wise \texttt{ValueSets}. After obtaining the \texttt{ValueSets} for each instruction, the \texttt{IdInstrTouchMem} procedure iterates over all instructions in a function and detects instructions that touch global or stack memory. Then it cross-references them with the previously generated \texttt{ValueSets} and the DWARF based $(\textit{Rgn}, \textit{offsets})$ to mark \BRel{}{} instructions and place those instruction addresses at \textit{instAddrs}.
Upon acquiring the result of \iVTR{}, we proceed to generate the complete ground truths to train the deep model of Task \task{3}, which is used to predict \BRel{} instructions for unknown binaries and to recover memory block boundaries with Algorithm~\ref{alg:value_tracking_analysis2}.

%% file: implementation.tex
\section{Implementation}\label{sec:implementation}

We implemented \sysname{} with PyTorch \cite{DBLP:conf/nips/PaszkeGMLBCKLGA19}.
We used Capstone 5.0.1~\cite{capstonepaper} to decode instructions. To accelerate this procedure, we used the C language API of Capstone \cite{capstone-c} to implement a multi-thread superset disassembly and instruction fields extraction procedure.
To implement \sysname{}'s \task{3}, our initial approach involved using Angr~\cite{angr} to develop \bVTR{} and \iVTR{} algorithms. However, we encountered limitations with Angr's default support for instruction-level granularity, leading us to modifying Angr's source code to obtain instruction-level value set information. However, this Angr-based implementation was  time- and resource-intensive; so we resorted to developing a Datalog-based lightweight value set analysis. This approach was combined with Python scripts to implement \bVTR{} and \iVTR{} algorithms.
To integrate \sysname{} into \bpa{}, we removed \bpa{}'s heuristic-based memory block generation step. We used the \bVTR{} procedure to convert the prediction results of \sysname{}'s \task{3} model into addressing values/offsets in specific memory regions. Such memory block boundaries were fed as input to the following steps of \bpa{}. Finally, we obtained the \sysname{}-facilitated \bpa{}, i.e., \bpaDisa{}.

We use 1,794 integers to encode the \textit{Opcode} field, including two integers for the Intel CET instructions \verb|ENDBR32| and \verb|ENDBR64|, and 768 integers for encoding VEX opcodes commonly seen in ICC binaries. \textit{ModRM} has 257 cases, in which one case represents the absence of \textit{ModRM}. Similarly, \textit{SIB} has 257 cases and one case represents \textit{SIB}'s absence. In our implementation, The length of input sequences is $L=512$, and the dimension of inputs to the transformer's encoders is $d_\text{model}=384$.
Our model has 6 encoder layers and 8 attention heads.
Compared with \xda{} \cite{DBLP:conf/ndss/PeiGWYJ21} (i.e., \#head$=12$, \#encoder-layer$=12$, $d_\text{model}=768$), our deep model is much simpler (Table~\ref{tab:hyperparameters}).
\sysname{}'s batch size is 16.
We observed class imbalance during the training procedure of our disassembly tasks, especially in the function entry-point identification.
Thus, we use the $\alpha$-balanced variant of Focal Loss \cite{DBLP:conf/iccv/LinGGHD17} as the loss function to ensure the training procedure focuses more on the sparse positive samples.

%% file: eval.tex
\section{Evaluation}\label{sec:evaluation}

Our evaluations aim to answer the following research questions:
\begin{compactenum}[RQ1.]
  \item How accurate is \sysname{} in identifying function entries and instructions compared with existing approaches?
  \item How robust is \sysname{} on unseen and obfuscated binaries?
  \item How effective is \sysname{} in deciding the memory block boundaries? How secure can \sysname{} achieve for \bpaDisa{}'s forward-edge resolving?
  \item How efficient is \sysname{} compared with existing approaches?
\end{compactenum}

\subsection{Dataset and Experimental Settings}

The benchmark binaries consist of non-obfuscated binaries \SNall{} and \SPall{}, real-world binaries \SRall{}, and obfuscated binaries \SOall{}, as presented in Table~\ref{tab:benchmark}. \SNall{}, \SRall{}, and \SOall{} are for tasks \task{1} and \task{2}, while \SPall{} is for task \task{3}.
We classify the benchmarks in \SNall{} based on the file format and ISA, e.g., \SN{ELF}{x86}.
The non-obfuscated binaries are from BAP corpora \cite{DBLP:conf/cav/BrumleyJAS11}, LLVM11 for Windows\footnote{In this dataset, we use the binaries smaller than 5 MB. URL: \url{https://drive.google.com/file/d/1UfS4YsbKWw6Xlp7NXf4tTHDN7gzDRY7p}}, or built from SPEC CPU 2006 and 2017 benchmarks with different compilers (GCC, Clang, ICC, and MSVC).
We build the SPEC CPU benchmark binaries into x86 and x64 architectures on different optimization levels, i.e., \OL{0}-\OL{3} and \OL{s} for GCC, Clang, and ICC; \OL{d}, \OL{1}, \OL{2}, and \OL{x} for MSVC. We deduplicate the length-$L$ input sequences in each sub-dataset of \SNall{} to avoid the \emph{train-test overlap}~\cite{DBLP:conf/ndss/PeiGWYJ21}.
The numbers of input sequences after deduplication are in Table~\ref{tab:benchmark}.
\SRall{} comprises 76 real-world non-obfuscated binaries of 10 applications from \cite{DBLP:conf/uss/YuQHY22}, one shared object file Glibc-2.22 from \cite{DBLP:conf/uss/AndriesseCVSB16}, and another subset \SR{ELF}{BT}. \SR{ELF}{BT} has seven \OL{1}-\OL{3} binaries built with extremely complex optimization options obtained by setting the \OL{0}-binary as BinTuner's baseline.
We build the obfuscated x64 binaries of \SOall{} on \OL{0}-\OL{3} with \ollvmfull{} \cite{ieeespro2015-JunodRWM, ollvm-repo} (\ollvm{} for short), \desync{} \cite{DBLP:conf/wcre/KargenHWEHS22, desync-repo}, tigress \cite{tigress}, VMProtect~\cite{vmprotect}, and Themida~\cite{themida}.
We apply tigress on the source code of cryptographic algorithms (e.g., DST40, SHA-256, XTEA, COMP128, MD5) and use GCC on MinGW to cross-compile the obfuscated PE binaries. We obfuscate the crypto-algorithm PE binaries cross-compiled by GCC on MinGW using VMProtect and Themida.
We use the default option of \desync{}. We enable the \verb|-sub|, \verb|-fla|, \verb|-bcf|, and \verb|-split_num| options of \ollvm{} simultaneously\footnote{Our \ollvm{} version does not support the option \texttt{indibran} as enabled by \cite{DBLP:conf/uss/YuQHY22, DBLP:conf/ndss/PeiGWYJ21}.}. We use the options \verb|Flatten|, \verb|EncodeArithmetic|, and \verb|AddOpaque| of tigress simultaneously.
On average, the code section of the binary gets bloated by 5.78x with \ollvm{}, 2.08x with \desync{}, and 25.33x with tigress.
We use the binaries of binutils, coreutils, and MiBench \cite{mibench} (\SP{train}{x86}) to train \sysname{}'s \task{3} model and
the SPEC2k6 binaries on \OL{0}-\OL{3} as the testing dataset \SP{test}{x86} to evaluate \task{3}.

\begin{table}[!t]
\renewcommand{\arraystretch}{1.2}
\caption{Datasets}
\label{tab:benchmark}
\centering
\vspace{-3ex}
\resizebox{\columnwidth}{!}{
\begin{threeparttable}
\begin{tabular}{c|c|c|c|c|r}
\hline
\multicolumn{2}{c|}{Dataset} & Compiler & Origin & \#Binary & \#Inst Seq. \\
\hline
\multirow{9}*{\SNall{}} & \multirow{3}*{\SN{ELF}{x86}} & GCC-9.2, GCC-4.7.2 & SPEC2k6\&2017+BAP\tnote{*} & 731 & 510,527 \\
\cline{3-6}
& & Clang-14.0 & SPEC2k6\&2017 & 215 & 481,389 \\
\cline{3-6}
& & ICC-2021.6.0, ICC-14.0.1 & SPEC2k6\&2017+BAP\tnote{*} & 726 & 647,147 \\
\cline{2-6}

& \multirow{3}*{\SN{ELF}{x64}} & GCC-9.2, GCC-4.7.2 & SPEC2k6\&2017+BAP\tnote{*} & 816 & 665,778 \\
\cline{3-6}
& & Clang-14.0 & SPEC2k6\&2017 & 300 & 587,854 \\
\cline{3-6}
& & ICC-2021.6.0, ICC-14.0.1 & SPEC2k6\&2017+BAP\tnote{*} & 811 & 890,651 \\
\cline{2-6}

& \SN{PE}{x86} & MSVC-2022/2008 & SPEC2k6\&2017+LLVM11 & 341 & 1,093,236 \\
\cline{2-6}

& \SN{PE}{x64} & MSVC-2022/2008 & SPEC2k6\&2017+LLVM11 & 375 & 1,177,774 \\
\hline

\multirow{2}*{\SRall{}} & \multirow{2}*{\SR{ELF}{x64}} & GCC & \cite{DBLP:conf/uss/YuQHY22} & 76 & 68,838 \\
\cline{3-6}
 &  & GCC & Glibc-2.22 of \cite{DBLP:conf/uss/AndriesseCVSB16} & 1 & 2,187 \\
\cline{2-6}
 & \SR{ELF}{BT} & GCC-9.4 with BinTuner & SQLite, thttpd, bzip2, diffutils & 7 & 4,188 \\
\hline

\multirow{5}*{\SOall{}} & \SO{ELF}{DSY} & \desync{} on GCC-11.4 & SPEC2k6\&2017 & 99 & 361,611 \\ 
\cline{2-6}
& \SO{ELF}{OLL} & \ollvm{} on Clang-4.0.1 & SPEC2k6\&2017 & 100 & 867,894 \\ 
\cline{2-6}
& \SO{PE}{TGR} & tigress with GCC-14.1 & crypto-algorithms & 60 & 7,089 \\
\cline{2-6}
& \SO{PE}{VMP} & VMProtect with GCC-14.1 & crypto-algorithms & 16 & 151,156 \\
\cline{2-6}
& \SO{PE}{TMD} & Themida with GCC-14.1 & crypto-algorithms & 16 & 963 \\
\hline

\multirow{2}*{\SPall} & \SP{train}{x86} & GCC-7.5 & binutils, coreutils, \cite{mibench} & 2664 & 7,31,104 \\
\cline{2-6}
& \SP{test}{x86} & GCC-9.2 & SPEC2k6 & 36 & 39,232\\
\hline

\hline
\end{tabular}
\begin{tablenotes}
\footnotesize
\item[*] The binaries of BAP corpora are built with GCC-4.7.2 and ICC-14.0.1.
\end{tablenotes}
\end{threeparttable}
}
\vspace{-3ex}
\end{table}

\subsubsection{Ground Truths Preparation}

The ground truths of \SR{ELF}{x64} are from the related works \cite{DBLP:conf/uss/YuQHY22, DBLP:conf/uss/AndriesseCVSB16}. Here, we focus on the ground truths of \SNall{}, \SOall{}, and \SR{ELF}{BT}.
We follow \cite{DBLP:conf/uss/YuQHY22} to obtain the function entry ground truths of ELF binaries from the symbol table and symbol index using pyelftools \cite{pyelftools}.
The obfuscators on ELF do not strip the binaries after obfuscation, thus allowing this approach.
For the \SN{PE}{} binaries, we use Dia2dump \cite{dia2dump} to parse the PDB files and derive the function entries.
For \SO{PE}{TGR}, we use \verb|objdump| to obtain the function entry ground truths from the symbol table.
The function entry ground truths on \SO{PE}{VMP} and \SO{PE}{TMD} cannot be obtained without other deobfuscators; thus these datasets are used to evaluate \task{2}.
Like \cite{DBLP:conf/ndss/PeiGWYJ21}, we do not treat the thunks in PE binaries and the trampolines in the \verb|.plt| section of ELF binaries as function entries. We also remove them from the analysis results to avoid mistaken false positives.

To obtain the instruction ground truth, we use linear disassembly with Capstone to obtain the ground truth of the GCC/Clang-built \SN{ELF}{} and \SR{ELF}{BT} binaries because, according to \cite{DBLP:conf/uss/AndriesseCVSB16}, GCC and Clang never inline data into the code section.
Overcoming the inaccuracy of Capstone's per-instruction decoding is out of our scope. Capstone is a rational choice considering it is used by other frameworks, e.g., \cite{triton, DBLP:conf/ndss/PeiGWYJ21, DBLP:conf/sp/YeZSAZ23}. One can transfer to other decoders, e.g., \cite{zydis, ida-pro}, as needed.
ICC and MSVC introduce data, e.g., jump tables, into the code section.
For the ICC-built \SN{ELF}{} binaries, we use \ida{}'s API to locate the data slots in the code section. We manually confirm the slots that indeed hold data. Then, we exclude the data and use linear disassembly on the rest code sub-sections to obtain the true instructions.
The parsing results of Dia2dump on \SN{PE}{} binaries contain the data addresses in the code section and labels. We follow \cite{DBLP:conf/uss/YuQHY22} to exclude the data slots from the code section to obtain the ground truth.
We treat \verb|NOP| in ELF and \verb|INT3| in PE binaries as padding and do not count them as positive or negative.

The obfuscators are built on different compilers; thus, investigating the compiler-backend emitted assembly requires much effort. In contrast,
we use a dynamic approach to obtain the sound but incomplete ground truths for the obfuscated binaries in \SOall{}. We developed a Pin tool to instrument the \SOall{} binaries and record the address and size of each runtime-reached instruction under standard workloads or typical test cases.
Because the true instruction segments are separated by \emph{unknown} slots that are unreached at runtime, we treat a predicted instruction as positive if it overlaps at least one byte with a true instruction. Otherwise, this prediction is classified as unknown.
We observed that over 50\% of runtime-reached instructions of \SO{PE}{VMP} binaries are in section \texttt{.vmp1}, while in \SO{PE}{TMD} binaries, over 98\% of the runtime-reached instructions are in section \texttt{.themida}. These sections contain the encrypted/compressed data or code that the static disassemblers cannot disassemble. We deem it a limitation of static disassemblers. As a result, in Section~\ref{subsec:generalizability}, we investigate the static disassemblers over section \texttt{.vmp2} of \SO{PE}{VMP} binaries and section \texttt{.text} of \SO{PE}{TMD} binaries, which hold the static machine code whose ground truths can be obtained by our Pin tool.

To obtain the true memory block boundaries on \SP{train}{x86} and \SP{test}{x86} for \task{3}, we follow the process in Section~\ref{subsubsec:gt-train}.
We use pyelftools to extract the binary's DWARF information and derive the raw memory block boundaries in both global sections and stack frames.
The DWARF information is not directly suitable for our purposes, since we encounter location information missing particularly in cases where the location information includes complex DWARF stack details, such as pointer deference, which cannot be resolved statically.
Following the DWARF-based memory block extraction, we utilize the \iVTR{} analysis in Section~\ref{subsubsec:gt-train} to obtain the ground truth \BRel{} instructions. The labeled instructions extracted from \SP{train}{x86} are then employed to train the \sysname{} deep model for \task{3}, while the ground truth \BRel{} instructions from \SP{test}{x86} are utilized for evaluations in Section~\ref{subsec:bpa-eval}.

\subsection{Baselines}\label{subsec:baseline}

To answer RQ1, RQ2, and RQ4, we use the learning-based approaches, \xda{}~\cite{DBLP:conf/ndss/PeiGWYJ21}, \birnn{}~\cite{DBLP:conf/uss/ShinSM15}, and \deepdi{}~\cite{DBLP:conf/uss/YuQHY22} as the baselines.
We train the \birnn{} deep models and finetune the \xda{} deep models using our datasets (more details in Appendix~\ref{app:baseline-impl}).
We have yet to obtain \deepdi{}'s implementation to train the deep model but can only access the released \deepdi{} model through its APIs. We call such a deep model the reference model \deepdi$_\textsf{ref}$ in Table~\ref{tab:deep-models}.
Table~\ref{tab:deep-models} denotes the deep models with the \emph{approach name} followed by the \emph{training dataset}.
Specifically, \xdaModel{\SNall}, \birnnModel{\SNall}, and \disaModel{\SNall} are the deep models trained/finetuned by the randomly selected 90\% of dataset \SNall. The rest 10\% of \SNall{}, i.e., \SN{test}{}, is the testing dataset.
The models of different tasks (\task{1} and \task{2}) are trained with different ground truths; thus, for example, \disaModel{\SNall} has two deep models, one for task \task{1} and the other for task \task{2}.
These models are the most general models we can obtain to compare with \deepdi's reference model on accuracy and generalizability. Moreover, we train more deep models with the dataset \SN{ELF}{x64}, called \emph{x64ELF-centric} models, to further investigate the generalizability, despite their incomparability with \deepdi$_\textsf{ref}$. We also train optimization-level-specific deep models, e.g., \sysname{}-\SNsub{ELF}{x64}{0}, standing for the \sysname{} model trained with the \OL{0}-binaries of \SN{ELF}{x64}.
Besides the deep models, We also use the popular reverse-engineering frameworks \ida{} 7.6, \ghidra{} 10.3, and \ninja{} 4.0 under their default settings as baselines. To make fair comparisons with these frameworks, we input the stripped binaries to these frameworks in our testing.
The metrics of RQ1 and RQ2 are the \emph{precision} (P), \emph{recall} (R), and F1-score.

\begin{table}[!t]
\renewcommand{\arraystretch}{1.2}
\caption{Deep Models for Evaluations of RQ1 and RQ2}
\label{tab:deep-models}
\centering
\vspace{-3ex}
\resizebox{\columnwidth}{!}{
\begin{threeparttable}

\begin{tabular}{c|c|c|c|c|c}
\hline
 & Testing Dataset for Accuracy & \multicolumn{4}{|c}{Testing Dataset for Generalizability} \\
 & \SN{test}{}\tnote{2} & \SRall & \multicolumn{2}{|c|}{\SOall} & \SN{PE}{} \\
\hline

 & \xdaModel{\SNall} & \xdaModel{\SNall} & \xdaModel{\SNall} & \xdaModel{\SN{ELF}{x64}} & \xdaModel{\SN{ELF}{x64}}\\

Model & \birnnModel{\SNall} & \birnnModel{\SNall} & \birnnModel{\SNall} & \birnnModel{\SN{ELF}{x64}} & \birnnModel{\SN{ELF}{x64}} \\

Name & \deepdi{}$_\textsf{ref}$ & \deepdi{}$_\textsf{ref}$ & \deepdi{}$_\textsf{ref}$ & N/A\tnote{1} & N/A\tnote{1} \\

 & \disaModel{\SNall} & \disaModel{\SNall} & \disaModel{\SNall} & \disaModel{\SN{ELF}{x64}} & \disaModel{\SN{ELF}{x64}} \\
\hline
\end{tabular}
\begin{tablenotes}
\item[1] \deepdi's reference model has the knowledge of x86 and PE binaries, thus we cannot compare its generalizability with the deep models trained only with the x64 ELF binaries (i.e., \xdaModel{\SN{ELF}{x64}}/\birnnModel{\SN{ELF}{x64}}/\disaModel{\SN{ELF}{x64}}).
\item[2] \deepdi's reference model has no knowledge of ICC-built binaries, thus in the testing dataset for the accuracy of \deepdi, we exclude the ICC-built binaries from the ground truths.
\end{tablenotes}
\end{threeparttable}
}
\vspace{-3ex}
\end{table}

To answer RQ3, our baselines are \bpa{} and \bpaDwarf{}. Similar to \bpaDisa{}, the \bpaDwarf{} is another modified version of \bpa{}, which utilizes DWARF-based memory blocks instead of \bpa{}'s heuristics-based memory blocks.
The memory blocks based on DWARF use debugging information to create the most precise memory blocks. Therefore, for analysis methods like \bpa{} that rely on memory block-based aliasing analysis, these nearly perfect memory blocks are expected to produce the best possible indirect call target result. Thus, \bpaDwarf{} represents the most achievable reference version of BPA.
To address RQ3's first question, we assess the effectiveness of memory blocks generated by \sysname{} compared to those from \bpa{}, using DWARF-based memory blocks as the ground truth reference.
To define the \emph{precision}, \emph{recall}, and F1-score in this scenario, a true positive (TP) indicates a correctly generated memory block that aligns with a block found in the reference ground truth. A false positive (FP) refers to a generated memory block that does not appear in the ground truth, while a false negative (FN) signifies a block present in the ground truth but absent from the memory block generation of \sysname{} or \bpa{}. To evaluate the second question of RQ3, we adopt the classic metric, AICT (Average Indirect Call Targets), to assess accuracy.
We also measure the recall to assess soundness, where a higher recall value indicates a smaller number of missed indirect call target predictions, which results in a more comprehensive CFG construction.
We leverage a Pin tool to gather runtime profile data for the \SP{test}{x86} benchmarks under the \verb|ref| workloads for recall checks.
Once we obtain the call-traces, following \bpa{}, we calculate the average recall for all indirect calls using
\begin{equation*}
\text{$Rc$} = \frac{1}{n} \sum_{i=1}^{n} \text{$Rc_i$} \quad \textit{where} \quad \text{$Rc_i$} = \frac{TP_i}{TP_i + FN_i}
\end{equation*}
\noindent Measuring precision is feasible. Considering that the precision can be estimated from a high recall and the AICT, we ignore this metric for simplicity.

\subsection{Accuracy (RQ1)}\label{subsec:accuracy}

\begin{figure}[!t]\centering
  \includegraphics[width=0.49\columnwidth]{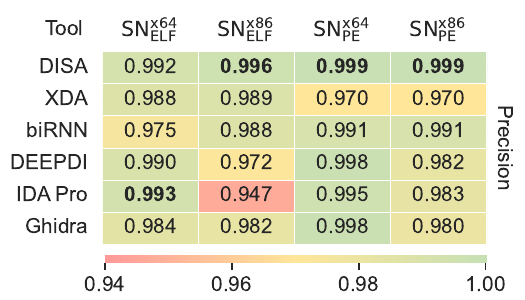}
  \hfill
  \includegraphics[width=0.49\columnwidth]{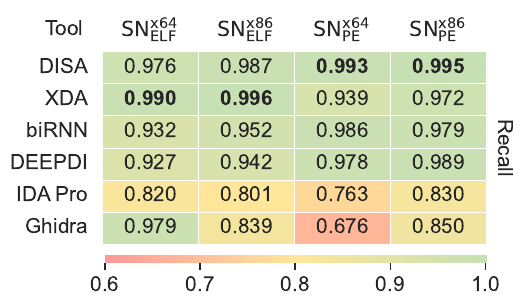}
\vspace{-2ex}
  \caption{\task{1} Accuracy Heatmaps on Different Platforms and ISA Variants}
  \label{fig:T1-accuracy}
\vspace{-3ex}
\end{figure}

We evaluate the accuracy of different learning-based approaches on four subsets of the testing dataset \SN{test}{}. We report the precision and recall on \task{1} in Fig.~\ref{fig:T1-accuracy}.
Because \deepdi's model training did not use the ICC-built binaries, we ignore the ground truths of the ICC-built binaries to decide \deepdi's accuracy.
Meanwhile, we also present \ida{} and \ghidra{}'s results on the complete four subsets of \SNall{}.
In predicting the function entries, \sysname{} has the best accuracy on \SN{PE}{}.
On \SN{ELF}{}, XDA has the best F1-score while \sysname{} has better precision than XDA.
The traditional frameworks allow more false negatives but are competitive in precision compared with the learning-based approaches.
On the PE binaries, the main reason is that we use the symbol table to collect the function boundary ground truth. The traditional frameworks, e.g., \ida{}, do not focus on resolving specific library function entries.

\subsection{Generalizability (RQ2)}\label{subsec:generalizability}

This section evaluates \sysname{}'s robustness. Specifically, we compare the deep models' generalizability in different scenarios, including on unseen real-world binaries and obfuscated binaries. We also discuss the cross-optimization-level generalizability of different deep models.

\subsubsection{Generalizability on Unseen Real-World Binaries}

\begin{figure}[!t]\centering
  \includegraphics[width=\columnwidth]{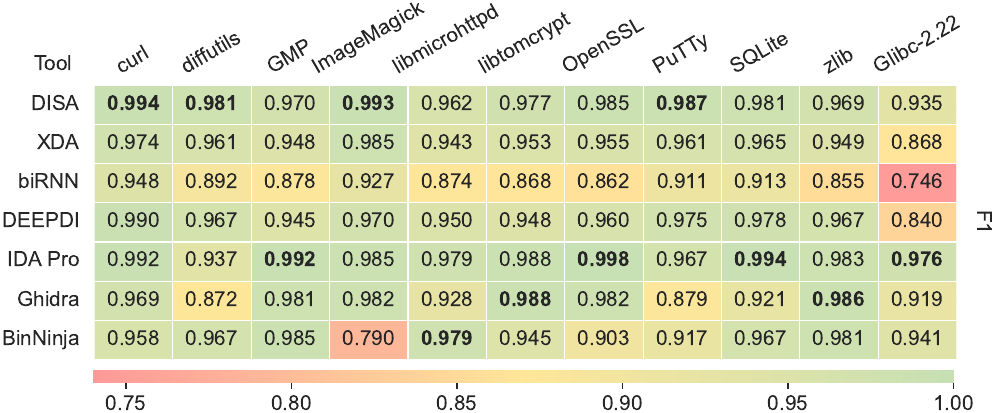}
\vspace{-4ex}
  \caption{\task{1} Generalizability Heatmap in Predicting Unseen Real-World x64-ELF Binaries}
  \label{fig:T1-generalizability}
\vspace{-2ex}
\end{figure}

We evaluate the robustness of different approaches on the real-world binaries in \SR{ELF}{x64}, as presented in Fig~\ref{fig:T1-generalizability}.
Generally, in task \task{1}, \sysname{} and \ida{} outperform other approaches.
\sysname{} shows a 1.8\% F1-score improvement over other deep-learning approaches in identifying function entries.
Specifically, the deep models' effectiveness on Glibc is not as good as on other programs due to its more complex function entries and the inline assembly code. In this case, \ida{} outperforms others.
In task \task{2}, the deep models all report F1$>$99\% (Fig.~\ref{fig:T2-generalizability-sr} in Appendix~\ref{app:eval-results}). \xda{} outperforms other deep models in identifying instructions of real-world binaries. We infer that \xda{}'s code-byte classification excels in more balanced classification. In contrast, with the high-level features, our superset-instruction classification could deal better with more imbalanced data for function entry-point identification.

\begin{figure}[!t]\centering
  \includegraphics[width=0.55\columnwidth]{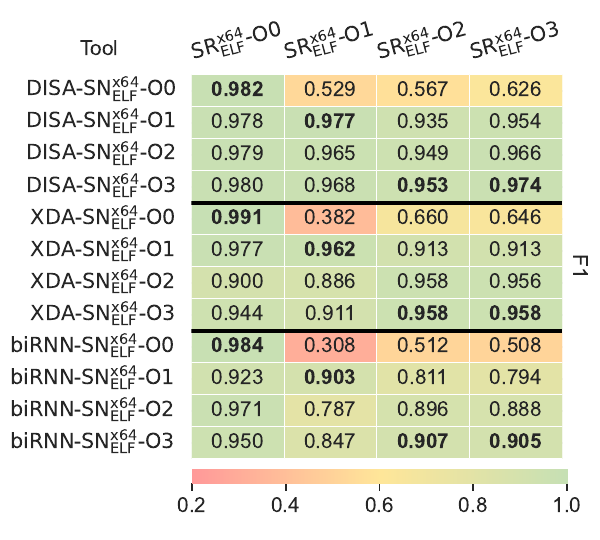}
\vspace{-2ex}
  \caption{\task{1} Cross-Optimization-Level Generalizability Heatmap on Real-World x64-ELF Binaries}
  \label{fig:T1-cross-opt-level}
\vspace{-3ex}
\end{figure}

\begin{figure}[!t]\centering
  \includegraphics[width=0.49\columnwidth]{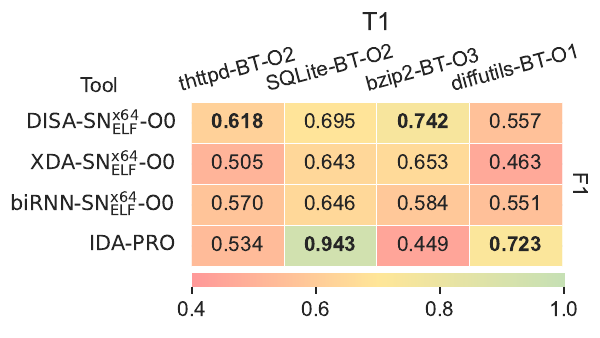}
  \hfill
  \includegraphics[width=0.49\columnwidth]{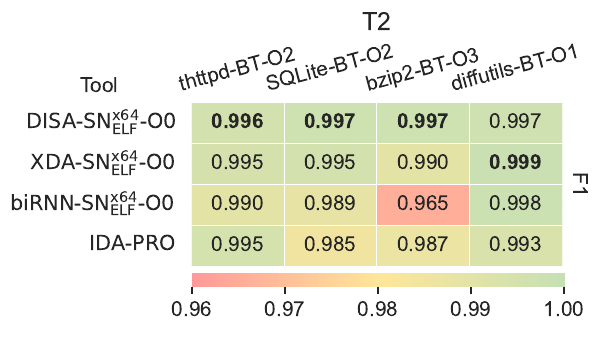}
\vspace{-2ex}
  \caption{Generalizability Heatmap on Real-World Binaries on BinTuner Discovered Complex Optimization Options}
  \label{fig:bintuner}
\vspace{-2ex}
\end{figure}

Unseen compiler optimization flags can introduce new instruction patterns and challenge the disassembly effectiveness. To this end, we first investigate the optimization-level-specific deep models' capability of cross-optimization-level predictions on real-world binaries.
From Fig.~\ref{fig:T1-cross-opt-level}, we know that for several optimization-level-specific models even not on \OL{0}, making predictions on \OL{0}-binaries is easier than on \OL{1}-\OL{3}. The \OL{0} models are extremely weak in predicting function entries of the binary on other optimization levels.
Then, we conduct another cross-optimization-level evaluation over the binaries built with the optimization flag combinations discovered by BinTuner. For a specific program, BinTuner can find the extremely complex optimization option for building the binary in maximum \emph{normalized compression distance} (NCD) with a baseline compilation configuration.
In Fig.~\ref{fig:bintuner}, we use four real-world programs, i.e., thttpd, SQLite, bzip2, and diffutils.
We use the \OL{0} binary of these programs as BinTuner's baseline for maximizing NCD.
The optimization flag combinations discovered by BinTuner are on \OL{1}-\OL{3}. Thus, for the cross-optimization-level evaluation, we use the \OL{0} deep models to decide the function entries and instructions. From Fig.~\ref{fig:bintuner}, we observe that BinTuner poses a significant challenge to the function entry-point identification but a minimal difficulty to the instruction disassembly. \sysname{} outperforms other deep models in most of these cases.

\subsubsection{Generalizability on Obfuscated Binaries}\label{subsubsec:obfuscated}

We compare the generalizability of the learning-based approaches in deciding the function entries and instructions of the obfuscated binaries of \SOall{}.
Apart from the best deep models of different approaches, we will also investigate whether the x64ELF-centric models could reach better performance.
Besides, we enable the compiler optimizations to investigate the impact of different optimizations on the obfuscation.

\begin{figure}[!t]\centering
  \includegraphics[width=\columnwidth]{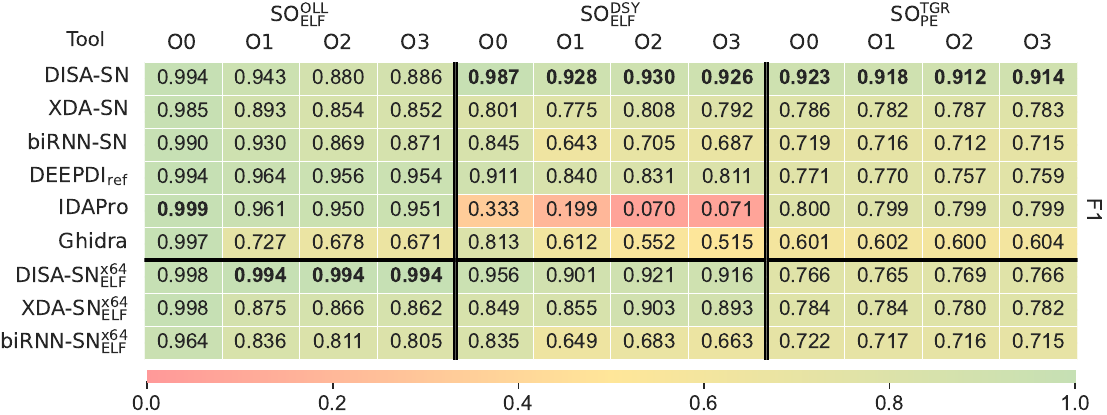}
\vspace{-4ex}
  \caption{\task{1} Generalizability Heatmap in Predicting Obfuscated Binaries at Different Optimization Levels}
  \label{fig:T1-generalizability-so}
\vspace{-3ex}
\end{figure}

Fig.~\ref{fig:T1-generalizability-so} presents the function entry-point identification results on obfuscated binaries.
The top half of Fig.~\ref{fig:T1-generalizability-so} presents the most general effect of different approaches. The bottom half presents the results of the x64ELF-centric models.
Generally, \ida{} and \ghidra{} have high precision while the deep models achieve better recall.
On \ollvm{}-obfuscated binaries (\SO{ELF}{OLL}), \ida{}, \deepdi{}, and \sysname{}'s x64ELF-centric model are the leading approaches.
On both the general and x64ELF-centric models, \sysname{} reports higher precision than \xda{}, while \xda{} enforces higher recalls.

Desync-cc and tigress have more intensive effects on disrupting  function entries than \ollvm{}.
On \SO{ELF}{DSY}, \sysname{}'s general model and x64ELF-centric model respectively outperform other approaches by 9.1\% and 5.7\% in F1-score. Therefore, \sysname{} has a significant advantage in detecting function entries of the binary obfuscated with the desynchronization technique.
To investigate the reason for \sysname{}'s advantage on \SO{ELF}{DSY}, we conduct case studies on several SPEC2017 binaries. On several functions starting with \texttt{mov}, \texttt{lea}, or logical \texttt{and} instruction, \xda{} and \deepdi{} have false negatives. \deepdi{} also ignores several function entries when the starting \texttt{push} instruction follows the \desync{}-inserted junk bytes.
The false positives of \xda{} and \deepdi{} are observed on several \desync{}-inserted \texttt{push} instructions and junk bytes. Generally, \deepdi{}'s GRU-model based function entry-point recovery only uses short instruction contexts and thus is less capable than \sysname{} in \task{1}. \sysname{}'s instruction-level function entry knowledge benefits the differentiation of \desync{}-inserted junk bytes compared to the byte-level knowledge used by \xda{}. Regarding \ida{} and \ghidra{}, they raise decoding errors on a considerable portion of instructions, thus failing to resolve function entries due to the missing instructions.

\begin{figure}[!t]\centering
  \includegraphics[width=0.5\columnwidth]{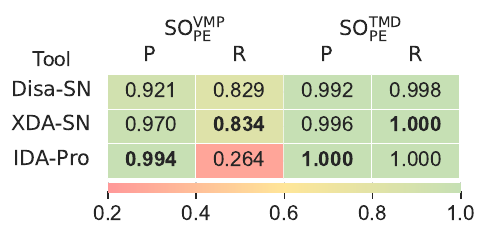}
\vspace{-2ex}
  \caption{\task{2} Generalizability Heatmap in Predicting VMProtect and Themida-obfuscated PE binaries}
  \label{fig:T2-generalizability-vmp-tmd}
\vspace{-4ex}
\end{figure}

On tigress-obfuscated binaries (\SO{PE}{TGR}), the model \disaModel{\SNall} outperforms other approaches by 13.2\% in F1-score, while \ida{} has the best precision. In general, on source-level obfuscators like \ollvm{} and tigress, the deep models do not have a significant advantage over the traditional frameworks, e.g., \ida{}.
In the function entry-point identification, we did not observe the interference from compiler optimizations to the obfuscation, which was assumed by \cite{DBLP:conf/ndss/PeiGWYJ21}, and the compiler optimizations are disabled by \cite{DBLP:conf/uss/YuQHY22, DBLP:conf/ndss/PeiGWYJ21}.
In many cases, the deep models have lower effectiveness on \OL{1}-\OL{3} binaries than on \OL{0} binaries, indicating that obfuscation combined with optimizations causes more intensive disruption to the pattern of function prologues.

Identifying instructions in obfuscated binaries is more straightforward than identifying function entries for the deep models (Fig.~\ref{fig:T2-generalizability-so} in Appendix~\ref{app:eval-results}). In task \task{2}, \ida{} is most effective on \ollvm{}-obfuscated binaries.
\deepdi{}$_\textsf{ref}$ and \xdaModel{\SNall} have the best instruction-level F1-score respectively on \desync{}-obfuscated and tigress-obfuscated binaries. When using the x64ELF-centric model, \disaModel{\SN{ELF}{x64}} outperforms others on \desync{}-obfuscated binaries.
In addition, we performed testing on the VMProtect and Themida-obfuscated binaries, i.e., \SO{PE}{VMP} and \SO{PE}{TMD}. As Fig.~\ref{fig:T2-generalizability-vmp-tmd} presented, the deep models outperform \ida{} in F1-score when predicting the instructions in the \texttt{.vmp2} section of VMProtect-obfuscated binaries. Disassembling the \texttt{.text} section of Themida-protected binary is easier for the tools to reach F1$\ge$99\%, while \ida{} has the best F1-score.

In summary, \sysname{} is more robust than other approaches to identify function entries, improving other deep-learning approaches by 1.8\% on unseen real-world binaries, 9.1\% on \desync{}-obfuscated binaries, and 13.2\% on tigress-obfuscated binaries.
Besides, \sysname{}'s x64ELF-centric model outperforms other approaches in identifying instructions of \desync{}-obfuscated binaries.

\subsection{Memory Block Boundary Determination and Security Effect (RQ3)}\label{subsec:bpa-eval}

We first compare \sysname{}'s memory block prediction effectiveness with \bpa{}'s original heuristic-based memory block boundary identification.
Then, we evaluate \bpaDisa{}'s accuracy in identifying indirect targets by comparing it with \bpa{} and \bpaDwarf{}.
Furthermore, after a case study on indirect call resolution, we conduct an ablation study to demonstrate each component's contribution to indirect call resolution.

\subsubsection{Effectiveness On Memory Block Boundary Generations}

We compared the memory block generation of \sysname{} with that of \bpa{}'s heuristic-based approach in terms of precision, recall, and F1-score defined in Section~\ref{subsec:baseline}.
This comparison allows us to assess the effectiveness of \sysname{} in minimizing false positive predictions and enhancing precision, which is vital for downstream applications such as indirect target identification.

\begin{figure}[!t]\centering
  \includegraphics[width=0.9\columnwidth]{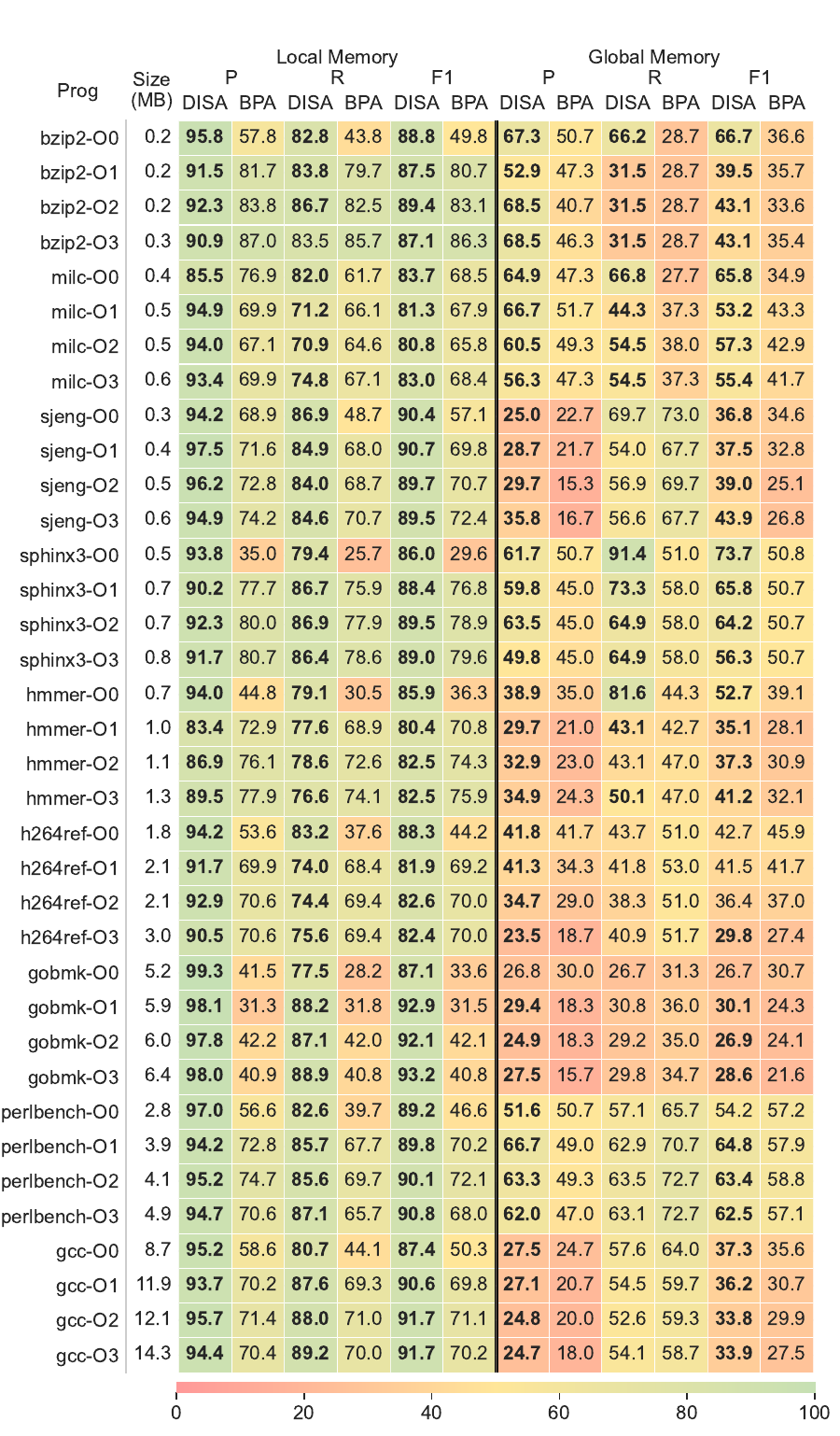}
\vspace{-2ex}
  \caption{Effectiveness Heatmap of \sysname{} on Deciding Memory Block Boundaries Compared with \bpa{} (in Percentage)}
  \label{fig:bpa-mem-block-effectiveness}
\vspace{-4ex}
\end{figure}

Compared to \bpa{}, \sysname{} demonstrates an average enhancement of 18.5\% in precision, 11.6\% in recall, and 16.1\% in F1-score when generating memory blocks. The prediction of memory blocks is further divided into local and global types. In local memory block partitioning, \sysname{} achieves an impressive 27\% and 24\% improvement in precision and F1-score respectively compared with \bpa{}, as presented in Fig.~\ref{fig:bpa-mem-block-effectiveness}. The local memory block prediction is uniformly good across increased binary complexity as well as different optimization levels. \sysname{} generally offers better local memory block boundaries than global memory.

Both \sysname{} and \bpa{} encounter significant challenges in accurately predicting global memory accesses. Our manual inspections on \SP{test}{x86} binaries reveal that global memory block start addresses are seldom loaded directly into registers; rather, they are typically accessed indirectly through complex pointer arithmetic. For example, in the dataset \SP{test}{x86}, 50\% of global memory blocks are accessed this way, with none of their starting addresses being explicitly loaded in the user code.
Accurately predicting them continues to pose challenges for both \sysname{} and \bpa{}. Furthermore, some compiler-allocated global memory blocks remain entirely unaccessed in user code, further complicating prediction accuracy. In addition,
\sysname{}'s intra-procedural value tracking fails to account for global accesses that are passed through function arguments.
Additionally, repetitive patterns such as loops can lead to false positives, as mentioned in Section~\ref{subsec:boundary-recovery}.

To overcome the global memory block prediction limitations, in \sysname{}, we have integrated the strengths of \task3{}'s machine learning-based predictions with BPA’s traditional static analysis. This hybrid approach has enhanced precision by leveraging \task3{}'s pattern identification alongside a modest increase in recall from \bpa{}'s conservative analysis, as illustrated in Fig.~\ref{fig:bpa-mem-block-effectiveness}.
On average, \sysname{} achieves a 10.1\% improvement in precision and an 8.1\% improvement in F1-score for global memory blocks compared with \bpa{}. Notably, \sysname{} has enhanced global memory block precision for most programs in \SP{test}{x86}. \sysname{} has surpassed \bpa{} in F1-score across all other binaries in \SP{test}{x86} except for h264ref. The primary reason for the performance dip in h264ref is also the complex pointer arithmetic involved in global block accessing.

Additionally, our evaluation demonstrates that \sysname{} improves the metrics at each optimization level.
As program complexity increases, \sysname{} exhibits more conservative memory block partitioning and outperforms \bpa{} at higher optimization levels, making it a more suitable tool for analyzing larger, real-world binaries.

\subsubsection{Indirect Call Target Comparison}

We evaluate the ability to resolve indirect call targets using \bpa{}'s block-boundary-based points-to analysis. The evaluation compares AICT values produced by original \bpa{}, \bpaDisa{}, and \bpaDwarf{}, as presented in Table~\ref{tab:bpa-effectiveness}.
For conciseness, we omit the results for bzip2, milc, sjeng, and sphinx3, as all three schemes
achieve identical AICT as well as 100\% recall.

\begin{table}[!t]
\renewcommand{\arraystretch}{1.2}\scriptsize
\caption{AICT and Recall Comparison between \bpaDisa{}, \bpaDwarf{}, and \bpa{} on \SP{test}{x86} (for GCC 9.2.0)}
\label{tab:bpa-effectiveness}
\centering
\vspace{-2ex}
\resizebox{\columnwidth}{!}{%
\begin{tabular}{c|c|c|c|c|c|c|c}
\hline
\multirow{2}{*}{\textbf{Program}} & \multirow{2}{*}{\textbf{\#iCallsite}} & \multicolumn{3}{c|}{\textbf{AICT}} & \multicolumn{3}{c}{\textbf{Recall}} \\ \cline{3-8}
                                  &                                       & \textbf{\bpaDwarf{}} & \textbf{\bpa{}} & \textbf{\bpaDisa{}} & \textbf{\bpaDwarf{}} & \textbf{\bpa{}} & \textbf{\bpaDisa{}} \\ \hline

hmmer-\OL{0}   & 9   & \textbf{2.8}  & 2.9  & \textbf{2.8}  & 100 & 100 & 100 \\
hmmer-\OL{1}   & 11  & 4.3           & 4.3  & 4.3           & 100 & 100 & 100 \\
hmmer-\OL{2}   & 10  & \textbf{2.7}  & 2.8  & 2.8           & 100 & 100 & 100 \\
hmmer-\OL{3}   & 9   & 1.0           & 1.0  & 1.0           & 100 & 100 & 100 \\ \hline

h264ref-\OL{0} & 369 & \textbf{4.2}  & 5.7  & \textbf{4.3}  & 100 & 100 & 100 \\
h264ref-\OL{1} & 353 & \textbf{4.1}  & 5.2  & 5.2           & 99.7 & 99.7 & 99.7 \\
h264ref-\OL{2} & 352 & \textbf{26.4} & 26.7 & 26.9          & 100 & 100 & 100 \\
h264ref-\OL{3} & 355 & \textbf{18.0} & 18.5 & 18.6          & 99.7 & 99.7 & 100 \\ \hline

gobmk-\OL{0}   & 44  & \textbf{846.9}  & 884.6  & \textbf{846.9}  & 100 & 100 & 100 \\
gobmk-\OL{1}   & 44  & \textbf{1334.7} & 1336.3 & 1336.3         & 100 & 100 & 100 \\
gobmk-\OL{2}   & 44  & 1337.7          & 1337.7 & 1337.7         & 100 & 100 & 100 \\
gobmk-\OL{3}   & 44  & 1416.2          & 1416.2 & 1416.2         & 100 & 100 & 100 \\ \hline

perlbench-\OL{0} & 139 & \textbf{388.9} & 400.3 & \textbf{387.6} & 100 & 100 & 99.1 \\
perlbench-\OL{1} & 139 & \textbf{370.1} & 379.7 & 379.7          & 100 & 100 & 100 \\
perlbench-\OL{2} & 110 & \textbf{371.2} & 377.6 & \textbf{373.9}          & 100 & 100 & 100 \\
perlbench-\OL{3} & 237 & 453.4          & 453.4 & 456.4          & 100 & 100 & 100 \\ \hline

gcc-\OL{0}     & 459 & \textbf{89.4}   & 540.8 & \textbf{447.1} & 95.4 & 99.3 & 95.3 \\
gcc-\OL{1}     & 473 & \textbf{336.1}  & 496.9 & \textbf{338.1} & 96.0 & 99.5 & 95.8 \\
gcc-\OL{2}     & 450 & \textbf{414.7}  & 440.9 & \textbf{431.1}    & 96.6 & 99.5 & 96.2 \\
gcc-\OL{3}     & 727 & \textbf{441.6}  & 526.7 & \textbf{454.1} & 97.8 & 99.6 & 96.2 \\

\hline
\end{tabular}%
}
\vspace{-3ex}
\end{table}

Table~\ref{tab:bpa-effectiveness} supports \sysname{}'s effectiveness in improving the indirect call targets prediction and substantiates our hypothesis about \bpaDwarf{}'s ultimate performance in Section~\ref{subsec:baseline}.
Compared to \bpaDwarf{}, \bpa{} shows deficiency in 15 out of 36 benchmarks in terms of AICT,
while \bpaDisa{} matches \bpaDwarf{}'s performance in 9 instances, indicating its superiority over \bpa{}.
\sysname{}'s advantage on memory block generation in Fig.~\ref{fig:bpa-mem-block-effectiveness} leads to noticeable AICT decreases for \OL{0} binaries and the more complex binaries in Table~\ref{tab:bpa-effectiveness}.
As program complexity increases, \bpa{}'s heuristic-based memory-block generation becomes limited, while \bpaDisa{} demonstrates better results, e.g., for benchmarks gobmk-\OL{0}, perlbench-\OL{0}, and gcc.
In profiling-based evaluation, we observed good recalls of \bpaDisa{} and \bpa{} in alignment with \bpaDwarf{} and a better precision of \bpaDisa{} for larger binaries. In addition, for h264ref-\OL{3}, we observe increased recall rates, indicating more coverage of the indirect call targets with increased AICT.
Furthermore, We noted a slight decrease in AICT for perlbench-\OL{0} compared with \bpaDwarf{}, attributed to missing targets and reduced recall rates.  Additionally, perlbench-\OL{3} and h264ref-\OL{2} had an indirect call prediction that neither \bpa{} nor \bpaDwarf{} reported. This observation indicates that \sysname{}'s higher recall in this case in Fig.~\ref{fig:bpa-mem-block-effectiveness} cannot avoid a false negative prediction on specific memory block boundary to introduce certain false indirect targets.
In addition, we identify a discrepancy between our reevaluation of \bpa{} and the original results presented in \cite{DBLP:conf/ndss/KimSZT21}.
Upon closer investigation, we confirm that an implementation issue mishandles finer-grained and coarse-grained memory blocks and causes this deviation.

\subsubsection{Case Study}

The accurate resolution of indirect function calls remains a significant challenge for commercial tools. In contrast, \bpaDisa{} delivers more precise CFGs than these tools. This case study examines the resolution of indirect calls within the bzip2 benchmark, using the source code for clarity. Notably, the same phenomenon can also be observed in the binary version of the code.

In bzip2, indirect calls specifically occur within two memory management functions: \texttt{default\_bzalloc} and \texttt{default\_bzfree}. Listing~\ref{lst:bzalloc} is a code snippet from bzip2 that defines a memory allocation structure and the corresponding default allocation functions.
The allocation and deallocation functions are set as defaults within the \texttt{BZ2\_bzCompressInit} function, as illustrated in Listing~\ref{lst:bzalloc2}. Additionally, indirect calls made via \texttt{BZALLOC} and \texttt{BZFREE} reference the functions \verb|default_bzalloc| and \texttt{default\_bzfree}, respectively.
These indirect calls are utilized in function \texttt{BZ2\_decompress}, which is detailed in Listing~\ref{lst:bzalloc4}. Specifically, function \texttt{BZ2\_decompress} invokes \texttt{default\_bzalloc} through \texttt{BZALLOC}.
Analysis with IDA and \ghidra{} shows that both tools fail to resolve the indirect calls to the function \texttt{default\_bzalloc}. In function \texttt{BZ2\_decompress}, they only identify direct calls to \verb|fprintf|, \verb|fwrite|, \verb|BZ2_indexIntoF|, \verb|makeMaps_d|, \verb|BZ2_hbCreateDecodeTables|, and \verb|BZ2_bz_AssertH_fail|.

In addition, our PIN tool-based detection framework effectively captures indirect calls to the \texttt{default\_bzalloc} function, confirming its reachability. This showcases \bpaDisa{}'s ability to create comprehensive CFGs. Our ability to resolve indirect call targets precisely is enabled by a practical block-based points-to analysis, where the precision of memory blocks is critical. \sysname{} \task{3} model helps us achieve this precision, resulting in clearer points-to sets and better resolution of indirect call targets.
While this study focuses on a small binary, we confirm that similar issues exist in larger binaries, further validating our method's superiority.

\begin{lstlisting}[language=C, float=h, caption=Memory allocation structure and associated calls, label={lst:bzalloc}, belowskip=-1.2\baselineskip]
typedef struct {
    void *(*bzalloc)(void *, int, int);
    void (*bzfree)(void *, void *);
    void *opaque;
} bz_stream;
static void* default_bzalloc(void* opaque, int items, int size) {
    return malloc(items * size);
}
static void default_bzfree(void* opaque, void* addr) {
    if (addr != NULL) free(addr);
}
\end{lstlisting}

\begin{lstlisting}[
    language=C, float=h, belowskip=-1.2\baselineskip,
    caption=Indirect initialization of calls,
    emph={BZALLOC, bzalloc, default_bzalloc},        % Highlight only BZALLOC
    emphstyle=\color{red}  % Apply red color to BZALLOC only
,label={lst:bzalloc2}][language=C, caption=Function Initialization]
int BZ_API(BZ2_bzCompressInit)(bz_stream* strm,_,_,_) {
    if (strm->bzalloc == NULL) strm->bzalloc = default_bzalloc;
    if (strm->bzfree == NULL) strm->bzfree = default_bzfree;
}
...
#define BZALLOC(nnn) (strm->bzalloc)(strm->opaque, (nnn), 1)
#define BZFREE(ppp) (strm->bzfree)(strm->opaque, (ppp))
\end{lstlisting}

\begin{lstlisting}[
    language=C, float=h,  belowskip=-0.4\baselineskip,
    caption=Indirect Calls in BZ2\_decompress,
    basicstyle=\footnotesize\ttfamily,
    emph={BZALLOC},        % Highlight only BZALLOC
    emphstyle=\color{red}  % Apply red color to BZALLOC only
,label={lst:bzalloc4}]
if (s->smallDecompress) {
    s->ll16 = BZALLOC(s->blockSize100k * num * sizeof(_));
    s->ll4  = BZALLOC(((1 + s->blockSize100k * num) >> 1) * sizeof(_));
    if (s->ll16 == NULL || s->ll4 == NULL) RETURN(BZ_MEM_ERROR);
} else {
    s->tt  = BZALLOC(s->blockSize100k * num * sizeof(_));
    if (s->tt == NULL) RETURN(BZ_MEM_ERROR);
}
\end{lstlisting}

\subsubsection{Ablation Study}
\label{subsubsec:ablation}

\begin{table}[!t]
\renewcommand{\arraystretch}{1.4}
\caption{AICT and Recall Comparison between \bpaDisa{}, BPA$_{\text{\sysname{}}}^{\text{T1}}$, BPA$_{\text{\sysname{}}}^{\text{T2}}$, and BPA$_{\text{\sysname{}}}^{\text{T12}}$ on \SP{test}{x86} (for GCC 9.2.0)}
\label{tab:bpa-ablation}
\centering
\vspace{-2ex}
\resizebox{\columnwidth}{!}{%
\begin{tabular}{c|c|c|c|c|c|c|c|c}
\hline
\multirow{2}*{\textbf{Program}} & \multicolumn{4}{c|}{\textbf{AICT}} & \multicolumn{4}{c}{\textbf{Recall}} \\ \cline{2-9}
                 & \textbf{\bpaDisa{}} & \textbf{BPA$_{\text{\sysname{}}}^{\text{T1}}$} & \textbf{BPA$_{\text{\sysname{}}}^{\text{T2}}$} & \textbf{BPA$_{\text{\sysname{}}}^{\text{T12}}$}
                 & \textbf{\bpaDisa{}} & \textbf{BPA$_{\text{\sysname{}}}^{\text{T1}}$} & \textbf{BPA$_{\text{\sysname{}}}^{\text{T2}}$} & \textbf{BPA$_{\text{\sysname{}}}^{\text{T12}}$} \\
\hline

milc-\OL{0} & 2.0 & 2.0 & 1.0 & 1.0 & 100.0 & 100.0 & 50.0 & 50.0 \\
milc-\OL{1} & 2.0 & 2.0 & 2.0 & 2.0 & 100.0 & 100.0 & 100.0 & 100.0 \\  
milc-\OL{2} & 2.0 & 2.0 & 2.0 & 2.0 & 100.0 & 100.0 & 100.0 & 100.0 \\  
milc-\OL{3} & 1.0 & 0.8 & 1.0 & 0.8 & 100.0 & 75.0 & 100.0 & 75.0 \\
\hline
sjeng-\OL{0} & 7.0 & 7.0 & 7.0 & 7.0 & 100.0 & 100.0 & 100.0 & 100.0 \\
sjeng-\OL{1} & 7.0 & \textbf{6.0} & 7.0 & \textbf{6.0} & 100.0 & 100.0 & 100.0 & 100.0 \\
sjeng-\OL{2} & 7.0 & 7.0 & 7.0 & 7.0 & 100.0 & 100.0 & 100.0 & 100.0 \\
sjeng-\OL{3} & 7.0 & 7.0 & 7.0 & 7.0 & 100.0 & 100.0 & 100.0 & 100.0 \\
\hline

h264ref-\OL{0} & 4.3 & 4.3 & 4.3 & 4.3 & 100.0 & 100.0 & 100.0 & 100.0 \\ 
h264ref-\OL{1} & 5.2 & 5.2 & 5.2 & 5.2 & 99.7 & 99.7 & 99.7 & 99.7 \\ 
h264ref-\OL{2} & 26.9 & 26.9 & 26.9 & 26.9 & 100.0 & 100.0 & 100.0 & 100.0 \\ 
h264ref-\OL{3} & 18.6 & \textbf{17.9} & 18.6 & \textbf{17.9} & 100.0 & 100.0 & 100.0 & 100.0 \\
\hline
gobmk-\OL{0} & 846.9 & 846.9 & 846.9 & 846.9 & 100.0 & 100.0 & 100.0 & 100.0 \\
gobmk-\OL{1} & 1336.3 & \textbf{1333.3} & 1336.3 & \textbf{1333.3} & 100.0 & 99.7 & 100.0 & 99.7 \\
gobmk-\OL{2} & 1337.7 &  \textbf{1330.3} & 1337.7 &  \textbf{1330.3} & 100.0 & 100.0 & 100.0 & 100.0 \\
gobmk-\OL{3} & 1416.2 & \textbf{1408.3} & 1416.2 & \textbf{1408.3} & 100.0 & 100.0 & 100.0 & 100.0 \\
\hline
perlbench-\OL{0} & 387.6 & 387.8 & 387.8 & 387.8 & 99.1 & 99.1 & 99.1 & 99.1 \\
perlbench-\OL{1} & 379.7 & \textbf{371.3} & 380.0 &  \textbf{371.3} & 100.0 & 100.0 & 100.0 & 100.0 \\
perlbench-\OL{2} & 373.9 & \textbf{372.9} & 374.0 & \textbf{372.9} & 100.0 & 100.0 & 100.0 & 100.0 \\
perlbench-\OL{3} & 456.4 & 463.4 & 465.4 & 463.4 & 100.0 & 100.0 & 100.0 & 100.0 \\

\hline
\end{tabular}%
}
\vspace{-3ex}
\end{table}

To evaluate the contribution of each component of \sysname{}, we began by integrating \bpaDisa{} with \task{1}, resulting in BPA$_{\text{\sysname{}}}^{\text{T1}}$. Similarly, we replaced \bpaDisa{} instruction boundary with the boundary of \task{2}, yielding BPA$_{\text{\sysname{}}}^{\text{T2}}$. Finally, we combined all three tasks of \sysname{} with \bpaDisa{}, resulting in BPA$_{\text{\sysname{}}}^{\text{T12}}$.
The evaluation results are presented in Table~\ref{tab:bpa-ablation}. As shown, incorporating the three \sysname{}-supported tasks into BPA enhances the effectiveness of control flow graph recovery than only \task{3}-supported \sysname{}, i.e., original \bpaDisa{}, especially for larger and more highly optimized binaries like gobmk and perlbench. In addition, it maintains a high recall rate for the majority, indicating its overall capability of covering the attack surface. For conciseness, we omit the results for bzip2, hmmer, and sphinx3, as all four schemes achieve identical AICT as well as 100\% recall.
We excluded the GCC benchmark from the ablation study due to its significantly higher evaluation time and memory requirements.

\subsection{Efficiency (RQ4)}\label{app:efficiency}

\begin{figure}[!t]%
    \centering
    \subfloat[Efficiency on \desync{}-Obfuscated Binaries]{
        \includegraphics[width=0.85\columnwidth]{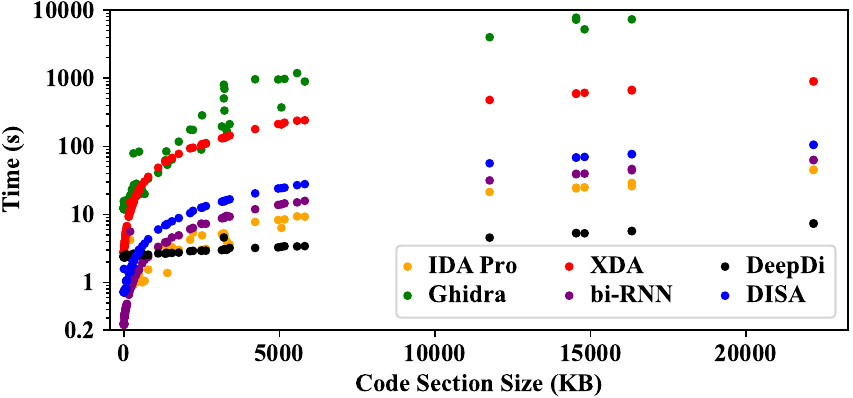}\label{subfig:efficiency-non-obfs}
        }\\
        \vspace{-2ex}
    \subfloat[Efficiency on \ollvm{}-Obfuscated Binaries]{
        \includegraphics[width=0.85\columnwidth]{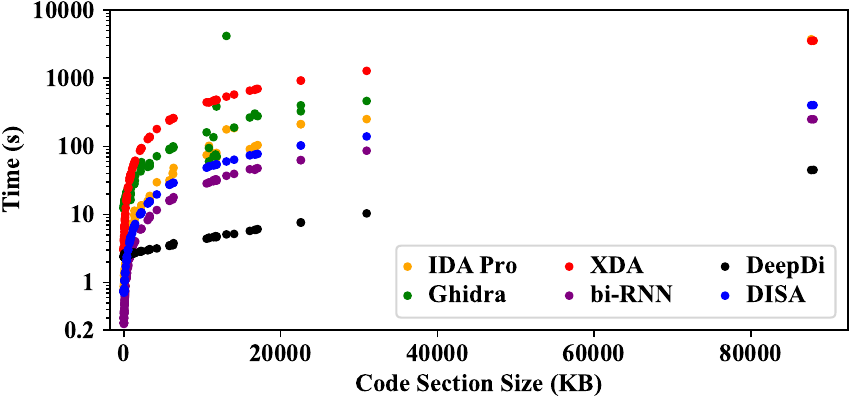}\label{subfig:efficiency-obfs}
        }
\vspace{-2ex}
    \caption{Task \task{2} Efficiency on Obfuscated Binaries}
    \label{fig:efficiency-supplementary}
\vspace{-2ex}
\end{figure}

We evaluate the efficiency of different approaches to identify instructions of the \SO{ELF}{DSY} and \SO{ELF}{OLL} binaries.
The prediction procedures of the learning-based approaches are on the NVIDIA A10 GPU, while \sysname{}'s multi-thread superset disassembly procedure is on the CPU.
The metric is the time cost from the binary input to all its code bytes or superset instructions getting classified completely.
According to the specific results in Fig.~\ref{fig:efficiency-supplementary}, the costs of learning-based approaches are stably correlated with the code section size.
Among the learning-based approaches, \deepdi{} is the most efficient due to the moderate model size and the GPU-accelerated superset disassembly.
\sysname{}'s multi-thread superset disassembly takes 3.55 seconds on average on each binary, which tends to be more costly than \deepdi{}.
\sysname{} is more costly than \birnn{} since \birnn{} has simpler models (Table~\ref{tab:hyperparameters} in Appendix~\ref{app:eval-results}).
\xda{} is more costly than \sysname{} in both prediction and model generation.
The costs of the traditional frameworks are irregular on different obfuscators.
\ida{} is more efficient on \desync{} than on \ollvm{} because on the \desync{}-obfuscated binaries, it fails to reach several portions of the code section.
\ghidra{} has extremely high costs on \desync{}-obfuscated binaries. \ghidra{} raises warnings about function body repairing on \desync{}-obfuscated binaries, indicating its inefficacy in handling desynchronization technique.
In summary, \sysname{} is more efficient than \xda{} and \ghidra{} and is competitive
with \ida{}, making it a scalable approach for real-world usage.

\section{Discussion}

Even though \sysname{}'s major goal is to enhance \bpa{} with improved memory blocks, we still compare our \bpaDwarf{} and \bpaDisa{} with the AICT metrics reported by the state-of-the-art indirect call prediction methods, CALLEE\cite{zhu2023callee} and AttnCall\cite{attncall}.
\bpaDwarf{} outperforms the others in 5 out of 9 benchmarks, i.e., milc, sphinx3, hmmer, h264ref, and gcc. Notably, for the largest benchmark, gcc, \bpaDwarf{} demonstrates a significant 67.4\% decrease in AICT compared with the state-of-the-art. \bpaDisa{} approaches \bpaDwarf{} in 4 out of those 5 benchmarks, showing potential in surpassing the state-of-the-art methods.

Besides, \sysname{} can be applied to other security applications that rely on memory blocks or similar concepts.
One possibility is to enhance binary fuzzing through hardware-assisted memory sanitizer \cite{DBLP:conf/uss/ChenSJL0DW023}. A critical part of the system is an object recovery scheme that progressively recovers object boundaries, resembling our memory blocks. The progressive object recovery relies on access patterns-based heuristics, which we propose to replace with our more precise memory blocks to enhance the binary fuzzing process.
It would also be an interesting future work to compare \sysname{}'s \task{1} and \task{2} capability with the LLM-based approach \cite{DBLP:journals/corr/abs-2407-08924}.

%% file: conclusion.tex
\section{Conclusion}\label{sec:conclusion}

\sysname{} is a new deep-learning-based disassembly approach based on the multi-head attention mechanism to classify superset instructions and achieve more precise boundary identifications.
\sysname{} disassembly framework uniquely supports predicting memory block boundaries.
We have integrated it with existing block memory model-based value set analysis to resolve the indirect call targets for the accurate binary-level CFG construction.
We have demonstrated that \sysname{} improves instruction and function entry-point identification, even for the binaries built with different obfuscators, and produces more accurate CFGs by establishing more precise memory block boundaries.

%% file: appendix.tex
\appendix

\section{Baseline Implementation and Configuration}\label{app:baseline-impl}

We re-implement \birnn{} using PyTorch following the same structure and setup described in \cite{DBLP:conf/uss/ShinSM15}. We especially follow the one-hot encoding of the instruction bytes and the \emph{rmsprop} optimization algorithms, which have yet to be fully adapted in the \birnn{} re-implementation provided by \xda{}.
We train the \birnn{} and \sysname{}'s deep models using our datasets.
Each \birnn{} model takes a batch size of 32.
\xda{} models rely on a pre-trained model.
In our evaluations, we use the pre-trained model provided by \xda{} and finetune the \xda{} models with our datasets.
Each epoch for finetuning an \xda{} model takes several days on an NVIDIA A10 GPU (24GB VRAM) with a 16-core vCPU (60GB RAM). In contrast, the training procedure of a \birnn{} model is the most efficient, and each epoch takes $<$2.5 hours and $<$10 hours for training the \disaModel{\SN{ELF}{x64}} and \disaModel{\SNall} models, respectively.
We use the default input sequence lengths $L_\text{\birnn{}}$=1000 and $L_\text{\xda{}}$=512.
For using the reference model of \deepdi{}, the \emph{slice length} of the testing dataset has a minor impact on the model's accuracy and generalizability. In our evaluations, we use the default slice length of 1024*1024.

\begin{figure}[t]\centering
  \includegraphics[width=0.6\columnwidth]{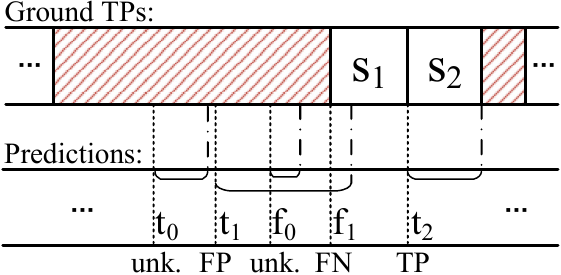}%
\vspace{-2ex}
  \caption{Classification Definition on Incomplete Ground Truths}
  \label{fig:gtp}
\vspace{-4ex}
\end{figure}

\paragraph{Instruction Ground Truth Collection on Obfuscated Binary}
We developed a Pin tool to instrument the \SOall{} binaries to record the address and size of each runtime-reached instruction.
In this procedure, we input the standard workloads of the SPEC benchmarks and typical test cases for the cryptographic algorithms.
Note that we do not use Pin's assembly translation to avoid the impact of its inaccuracy.
We calculate the \textit{coverage} of the runtime-reached instructions to the code section and exclude the binaries with \textit{coverage}$<$\%5. As a result, we obtained \SO{ELF}{DSY}, \SO{ELF}{OLL}, and \SO{PE}{TGR}.
The incomplete ground truths of obfuscated binaries require a new definition of classification. As illustrated in Fig.~\ref{fig:gtp}, the ground true instructions, e.g., s$_1$ and s$_2$, are spread in segments in the code section, and the instruction segments are separated by \emph{unknown} slots that are unreached at runtime.
The deep models predict on each code byte, e.g., the positive predictions t$_0$, t$_1$, and t$_2$, and the negative predictions f$_0$ and f$_1$.
If we disassemble an instruction falling into an unknown slot, e.g., on the byte of t$_0$ and f$_0$, we treat such bytes as unknown instead of positive or negative.
If the byte is predicted as an instruction boundary and this instruction has at least one-byte overlap with a true instruction, then this boundary byte has a positive classification.
For example, the positive verdict t$_1$ on a non-instruction byte is a false positive (FP). The negative verdict f$_1$ on the true instruction s$_1$ is a false negative (FN).
The positive verdict t$_2$ on the true instruction s$_2$ is a true positive (TP).

\section{Evaluation Results Supplementary}\label{app:eval-results}

\paragraph{Results on Accuracy}

\begin{figure}[!t]\centering
  \includegraphics[width=0.6\columnwidth]{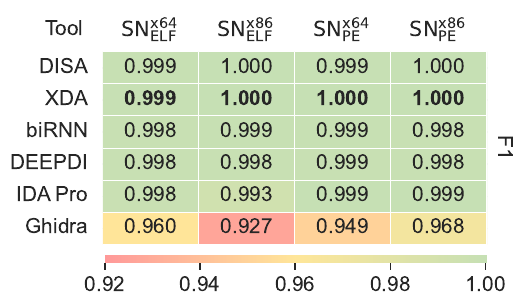}
\vspace{-2ex}
  \caption{T2 Accuracy Heatmaps on Different Platforms and ISA Variants}
  \label{fig:T2-accuracy}
\vspace{-3ex}
\end{figure}

The instruction-level deep models take sufficient knowledge of the instruction features on different platforms; thus, the \task{2} deep models have F1$>$99\%, and the comparative advantages between these models are generally insignificant (Fig.~\ref{fig:T2-accuracy}).

\begin{figure}[!t]\centering
  \includegraphics[width=\columnwidth]{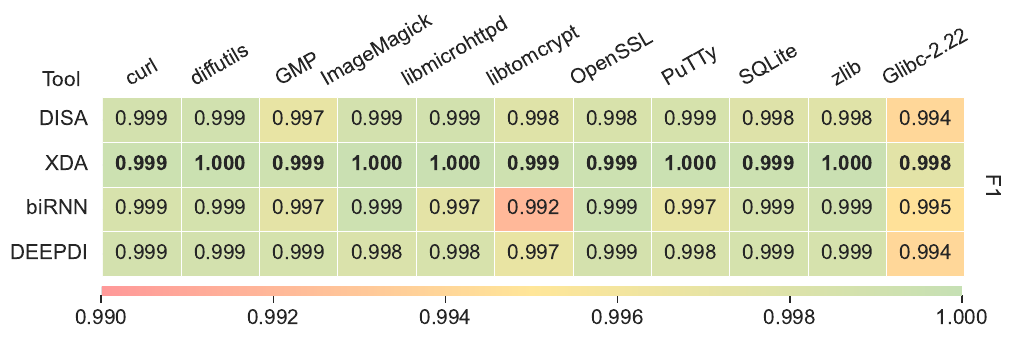}
\vspace{-4ex}
  \caption{\task{2} Generalizability Heatmap in Predicting Unseen Real-World x64-ELF Binaries}
  \label{fig:T2-generalizability-sr}
\vspace{-2ex}
\end{figure}

\begin{figure}[!t]\centering
  \includegraphics[width=\columnwidth]{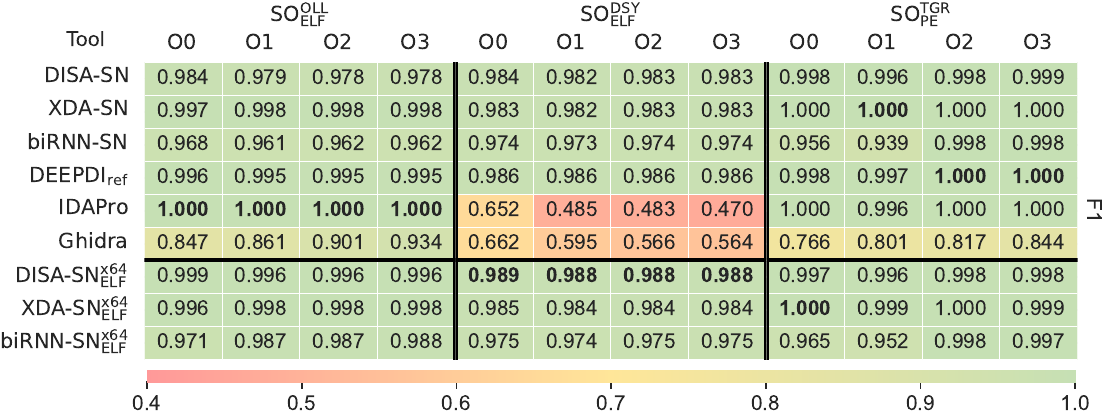}
\vspace{-3ex}
  \caption{\task{2} Generalizability Heatmap in Predicting Obfuscated Binaries at Different Optimization Levels}
  \label{fig:T2-generalizability-so}
\vspace{-2ex}
\end{figure}

\begin{table*}[!t]
\renewcommand{\arraystretch}{1.3}
\caption{Generalizability on Task \task{2} to Predict Obfuscated Binaries with Different Code Coverage in Ground Truth Collection}
\label{tab:generalizability-obfuscated-coverage}
\centering\scriptsize
\vspace{-3ex}
\begin{tabular}{c|c|c|c|c|c|c|c|c|c|c|c|c|c|c|c|c|c|c|c|c}
\hline
\multirow{4}*{Model}  & \multicolumn{8}{c|}{\SO{ELF}{OLL}} & \multicolumn{8}{c|}{\SO{ELF}{DSY}} & \multicolumn{4}{c}{\SO{PE}{TGR}} \\
\cline{2-21}
& \multicolumn{20}{c}{Code Coverage(\#Binaries)} \\

& \multicolumn{2}{c|}{5\%-30\%(25)} & \multicolumn{2}{c|}{30\%-50\%(31)} & \multicolumn{2}{c|}{50\%-75\%(39)} & \multicolumn{2}{c|}{75\%-100\%(5)} & \multicolumn{2}{c|}{5\%-30\%(31)} & \multicolumn{2}{c|}{30\%-50\%(26)} & \multicolumn{2}{c|}{50\%-75\%(33)} & \multicolumn{2}{c|}{75\%-100\%(9)} &
\multicolumn{2}{c|}{5\%-30\%(56)} & \multicolumn{2}{c}{30\%-50\%(4)} \\
\cline{2-21}
& P & R & P & R & P & R & P & R & P & R & P & R & P & R & P & R & P & R & P & R \\
\hline

\disaModel{\SN{ELF}{x64}} & \bl{.998} & \bl{.998} & .997 & \bl{.998} & .996 & .997 & .996 & .997 & \bl{.981} & \bl{.996} & \bl{.981} & \bl{.996} & \bl{.980} & \bl{.995} & \bl{.982} & \bl{.997} & .997 & .998 & .999 & .999 \\

\xdaModel{\SN{ELF}{x64}} & .996 & .995 & \bl{.998} & .997 & \bl{.998} & \bl{.997} & \bl{.998} & \bl{.997} & .976 & .992 & .977 & .992 & .976 & .991 & .978 & .993 & \bl{1.00} & \bl{.999} & \bl{1.00} & \bl{1.00} \\

\birnnModel{\SN{ELF}{x64}} & .979 & .969 & .981 & .979 & .985 & .984 & .986 & .986 & .972 & .978 & .972 & .977 & .970 & .975 & .970 & .979 & .966 & .981 & .998 & .998 \\
\hline
\end{tabular}
\vspace{-2ex}
\end{table*}

\paragraph{Results on Generalizability}
Fig.~\ref{fig:T2-generalizability-sr} illustrates the generalizability of deep models in identifying instructions of unseen real-world binaries in \SR{ELF}{x64}.
Fig.~\ref{fig:T2-generalizability-so} presents the results in identifying instructions of the binaries respectively obfuscated by \ollvm{}, \desync{}, and tigress. \sysname{}'s x64ELF-centric model outperforms other approaches to identify the instructions of the \desync{}-obfuscated binaries.
Due to the dynamic approach to obtaining the incomplete ground truths of \SOall{}, the classification errors (FPs and FNs) can only arise on the superset instructions that overlap with the runtime-reached true instructions. We further investigate how the code coverage of the ground true instructions impacts the detection results, as the higher code coverage implies fewer model predictions falling in the \emph{unknown} slots, and the detection results are closer to the results on complete ground truths.
We use the x64ELF-centric deep models.
Table~\ref{tab:generalizability-obfuscated-coverage} presents the results at different code coverages. Generally, we did not observe remarkable decreases in precision and recall along with increased code coverage of ground true instructions. The results justify our strategy of obtaining the incomplete ground truths at runtime.

\paragraph{Results on Efficiency}

\begin{table}[!t]
\renewcommand{\arraystretch}{1.3}\footnotesize
\caption{Number of Trainable Parameters of Deep Models}
\label{tab:hyperparameters}
\centering
\vspace{-2ex}
\begin{tabular}{c|c c c c}
\hline
 & \birnn{} & \deepdi{} & \sysname{} & \xda{} \\
\hline

\#Parameters & 8.8k & 49.9k & 12.9M & 86.8M \\
\hline
\end{tabular}
\end{table}

Besides comparing the time cost in Section~\ref{app:efficiency}, the prediction cost of different learning-based approaches can be estimated by comparing the number of trainable parameters of deep models, as presented in Table~\ref{tab:hyperparameters}.

\begin{table}[!t]
\renewcommand{\arraystretch}{1.5}  
\centering
\scriptsize  
\caption{Configurations of Different BPA Variants}
\label{tab:bpa-variants}
\vspace{-2ex}
\begin{tabular}{c|c|c|c|c|c|c|c|c}
\hline
\textbf{Configuration} & \multicolumn{3}{c|}{\textbf{\task{1}}} & \multicolumn{3}{c|}{\textbf{\task{2}}} & \multicolumn{2}{c}{\textbf{\task{3}}} \\
\textbf{Notation} & \textbf{IDA} & \textbf{\sysname{}} & \textbf{BPA} & \textbf{IDA} & \textbf{\sysname{}} & \textbf{BPA} & \textbf{\sysname{}} & \textbf{BPA} \\
\hline
\bpa{} &  &  & \CIRCLE &  &  & \CIRCLE &  & \CIRCLE \\

\bpaDisa{} &  &  & \CIRCLE &  &  & \CIRCLE & \CIRCLE &  \\

BPA$_{\text{\sysname{}}}^{\text{T1}}$ & & \CIRCLE & & & & \CIRCLE & \CIRCLE & \\

BPA$_{\text{\sysname{}}}^{\text{T2}}$ & & & \CIRCLE & & \CIRCLE & & \CIRCLE & \\

BPA$_{\text{\sysname{}}}^{\text{T12}}$ & & \CIRCLE & & & \CIRCLE & & \CIRCLE & \\

BPA$_{\text{\sysname{}}}^{\text{IDA1}}$ & \CIRCLE &  & &  &  & \CIRCLE & \CIRCLE &  \\

BPA$_{\text{\sysname{}}}^{\text{IDA2}}$ &  &  & \CIRCLE & \CIRCLE &  &  & \CIRCLE & \\

BPA$_{\text{\sysname{}}}^{\text{IDA12}}$ & \CIRCLE &  &  & \CIRCLE &  &  & \CIRCLE & \\

BPA$_{\text{\sysname{}}}^{\text{T1, IDA2}}$ &  & \CIRCLE &  & \CIRCLE &  &  & \CIRCLE &  \\

BPA$_{\text{\sysname{}}}^{\text{T2, IDA1}}$ & \CIRCLE &  &  &  & \CIRCLE &  & \CIRCLE &  \\
\hline
\end{tabular}%
\end{table}

\begin{table*}[!t]
\renewcommand{\arraystretch}{1.5}
\centering
\caption{AICT and Recall across Different BPA Variants}
\label{tab:ablation-aict_recall_summary}
\vspace{-1ex}
\resizebox{\textwidth}{!}{%
\begin{tabular}{c|c|c|c|c|c|c|c|c|c|c|c|c|c|c|c|c|c|c|c|c}
\hline
& \multicolumn{10}{c|}{\textbf{AICT}} & \multicolumn{10}{c}{\textbf{Recall}} \\ \hline
\textbf{Program}
& \texttt{\bpa{}}
& \texttt{\bpaDisa{}}
& BPA$_{\text{\sysname{}}}^{\text{T2}}$
& BPA$_{\text{\sysname{}}}^{\text{T1}}$
& BPA$_{\text{\sysname{}}}^{\text{T12}}$
& BPA$_{\text{\sysname{}}}^{\text{IDA12}}$
& BPA$_{\text{\sysname{}}}^{\text{IDA2}}$
& BPA$_{\text{\sysname{}}}^{\text{IDA1}}$
& BPA$_{\text{\sysname{}}}^{\text{T2, IDA1}}$
& BPA$_{\text{\sysname{}}}^{\text{T1, IDA2}}$
& \texttt{\bpa{}}
& \texttt{\bpaDisa{}}
& BPA$_{\text{\sysname{}}}^{\text{T2}}$
& BPA$_{\text{\sysname{}}}^{\text{T1}}$
& BPA$_{\text{\sysname{}}}^{\text{T12}}$
& BPA$_{\text{\sysname{}}}^{\text{IDA12}}$
& BPA$_{\text{\sysname{}}}^{\text{IDA2}}$
& BPA$_{\text{\sysname{}}}^{\text{IDA1}}$
& BPA$_{\text{\sysname{}}}^{\text{T2, IDA1}}$
& BPA$_{\text{\sysname{}}}^{\text{T1, IDA2}}$ \\
\hline
milc-\OL{0} & 2.0 & 2.0 & 1.0 & 2.0 & 1.0 & 2.0 & 2.0 & 2.0 & 1.0 & 2.0 & 100.0 & 100.0 & 50.0 & 100.0 & 50.0 & 100.0 & 100.0 & 100.0 & 50.0 & 100.0 \\

milc-\OL{3} & 1.0 & 1.0 & 1.0 & 0.8 & 0.8 & 1.0 & 1.0 & 1.0 & 1.0 & 0.8 & 100.0 & 100.0 & 100.0 & 75.0 & 75.0 & 100.0 & 100.0 & 100.0 & 100.0 & 75.0 \\
\hline

sjeng-\OL{1} & 7.0 & 7.0 & 7.0 & \textbf{6.0} & \textbf{6.0} & 7.0 & 7.0 & 7.0 & 7.0 & \textbf{6.0} & 100.0 & 100.0 & 100.0 & 100.0 & 100.0 & 100.0 & 100.0 & 100.0 & 100.0 & 100.0 \\

\hline
hmmer-\OL{0} & 2.9 & \textbf{2.8} & \textbf{2.8} & \textbf{2.8} & \textbf{2.8} & \textbf{2.8} & \textbf{2.8} & \textbf{2.8} & \textbf{2.8} & \textbf{2.8} & 100.0 & 100.0 & 100.0 & 100.0 & 100.0 & 100.0 & 100.0 & 100.0 & 100.0 & 100.0 \\

\hline
h264ref-\OL{0} & 5.7 & \textbf{4.3} & \textbf{4.3} & \textbf{4.3} & \textbf{4.3} & \textbf{4.3} & \textbf{4.3} & \textbf{4.3} & \textbf{4.3} & \textbf{4.3} & 100.0 & 100.0 & 100.0 & 100.0 & 100.0 & 100.0 & 100.0 & 100.0 & 100.0 & 100.0 \\
h264ref-O1 & \textbf{5.2} & \textbf{5.2} & \textbf{5.2} & \textbf{5.2} & \textbf{5.2} & \textbf{5.2} & \textbf{5.2} & \textbf{5.2} & \textbf{5.2} & \textbf{5.2} & 99.7 & 99.7 & 99.7 & 99.7 & 99.7 & 99.7 & 99.7 & 99.7 & 99.7 & 99.7 \\
h264ref-\OL{2} & \textbf{26.7} & 26.9 & 26.9 & 26.9 & 26.9 & 26.9 & 26.9 & 26.9 & 26.9 & 26.9 & 100.0 & 100.0 & 100.0 & 100.0 & 100.0 & 100.0 & 100.0 & 100.0 & 100.0 & 100.0 \\
h264ref-\OL{3} & 18.5 & 18.6 & 18.6 & \textbf{17.9} & \textbf{17.9} & 18.6 & 18.6 & 18.6 & 18.6 & 18.0 & 99.7 & 100.0 & 100.0 & 100.0 & 100.0 & 100.0 & 100.0 & 100.0 & 100.0 & 100.0 \\
\hline
gobmk-\OL{0} & 884.6 & \textbf{846.9} & \textbf{846.9} & \textbf{846.9} & \textbf{846.9} & \textbf{846.9} & \textbf{846.9} & \textbf{846.9} & \textbf{846.9} & \textbf{846.9} & 100.0 & 100.0 & 100.0 & 100.0 & 100.0 & 100.0 & 100.0 & 100.0 & 100.0 & 100.0 \\
gobmk-\OL{1} & 1336.3 & 1336.3 & 1336.3 & \textbf{1333.3} & \textbf{1333.3} & 1336.3 & 1336.3 & 1336.3 & 1336.3 & \textbf{1333.3} & 100.0 & 100.0 & 100.0 & 99.7 & 99.7 & 100.0 & 100.0 & 100.0 & 100.0 & 99.7 \\
gobmk-\OL{2} & 1337.7 & 1337.7 & 1337.7 & \textbf{1330.3} & \textbf{1330.3} & 1337.7 & 1337.7 & 1337.7 & 1337.8 & \textbf{1330.3} & 100.0 & 100.0 & 100.0 & 100.0 & 100.0 & 100.0 & 100.0 & 100.0 & 100.0 & 100.0 \\
gobmk-\OL{3} & 1416.2 & 1416.2 & 1416.2 & \textbf{1408.3} & \textbf{1408.3} & 1416.2 & 1416.2 & 1416.2 & 1416.2 & \textbf{1408.3} & 100.0 & 100.0 & 100.0 & 100.0 & 100.0 & 100.0 & 100.0 & 100.0 & 100.0 & 100.0 \\
\hline
perlbench-\OL{0} & 400.3 & \textbf{387.6} & 387.8 & 387.8 & 387.8 & \textbf{387.6} & \textbf{387.6} & \textbf{387.6} & 387.8 & 387.8 & 100.0 & 99.1 & 99.1 & 99.1 & 99.1 & 99.1 & 99.1 & 99.1 & 99.1 & 99.1 \\
perlbench-\OL{1} & 379.7 & 379.7 & 380.0 & \textbf{371.3} & \textbf{371.3} & 379.7 & 379.7 & 379.7 & 380.0 & \textbf{371.3} & 100.0 & 100.0 & 100.0 & 100.0 & 100.0 & 100.0 & 100.0 & 100.0 & 100.0 & 100.0 \\
perlbench-\OL{2} & 377.6 & 373.9 & 374.0 & \textbf{372.9} & \textbf{372.9} & 373.9 & 373.9 & 373.9 & 374.0 & \textbf{372.9} & 100.0 & 100.0 & 100.0 & 100.0 & 100.0 & 100.0 & 100.0 & 100.0 & 100.0 & 100.0 \\
perlbench-\OL{3} & \textbf{453.4} & 456.4 & 465.4 & 463.4 & 463.4 & 456.4 & 456.4 & 456.4 & 465.4 & 463.4 & 100.0 & 100.0 & 100.0 & 100.0 & 100.0 & 100.0 & 100.0 & 100.0 & 100.0 & 100.0 \\
\hline
\end{tabular}%
}
\end{table*}

\paragraph{Results on extended ablation study}
In our extended ablation study, we investigated a range of combinations of instruction and function boundaries derived from IDA, the original \bpa{}, and \sysname{}'s three tasks. As illustrated in Table~\ref{tab:bpa-variants}, we developed several \bpa{} variants by incorporating different instruction and function boundaries alongside a memory block boundary detection module either from \sysname{}'s \task{3} or from original \bpa{}. The performance outcomes for these variants, measured in terms of AICT and Recall, are presented in Table~\ref{tab:ablation-aict_recall_summary}. For clarity, we excluded bzip2, sphinx3, and a few optimized versions of milc, sjeng, hmmer, and h264ref, as all these instances yielded identical AICT results with a 100\% recall rate across all variants. Furthermore, we had to omit GCC from our study due to time constraints, as detailed in Section 5.5.4.
Table~\ref{tab:ablation-aict_recall_summary} demonstrates that our \bpa{} variants, particularly BPA$_{\text{\sysname{}}}^{\text{T12}}$, achieved results that are either comparable to or superior to other versions. The IDA-supported variants, including BPA$_{\text{\sysname{}}}^{\text{IDA12}}$, BPA$_{\text{\sysname{}}}^{\text{IDA2}}$, BPA$_{\text{\sysname{}}}^{\text{IDA1}}$, and BPA$_{\text{\sysname{}}}^{\text{T2, IDA1}}$, have demonstrated comparable performance, especially for lower optimized binaries,i.e., hmmer-\OL{0}, h264ref-\OL{0}, gobmk-\OL{0}, and perlbench-\OL{0}. The BPA$_{\text{\sysname{}}}^{\text{T1, IDA2}}$, which takes IDA instruction boundaries and \sysname{} function boundaries, has also matched the performance of BPA$_{\text{\sysname{}}}^{\text{T12}}$ for sjeng-\OL{1}, h264ref-\OL{0}, \OL{1}, hmmer-\OL{0}, gobmk, as well as perlbench-\OL{1} and \OL{2}. This suggests that \sysname{} \task{1} can enhance the performance of IDA-supported \bpa{} variants as well. Nonetheless, BPA$_{\text{\sysname{}}}^{\text{T12}}$ continues to outperform all others by delivering the best overall improvements.